# A Catalog of Automatically Identified Multi-Signature ICMEs Observed by Solar Orbiter

Felix N. Minta[12,3], Stefano A. Livi[3,2,4], George C. Ho[3,2], Susan T. Lepri[4], Heather A. Elliott[3,2]

## Abstract

In this study, we present a comprehensive catalog of 138 interplanetary coronal mass ejections (ICMEs) compiled using multiple distance-normalized ICME criteria within sliding time windows from Solar Orbiter in-situ measurements. We observed sub-adiabatic cooling in proton temperatures, a near-spherical radial decrease in density, a steep decline in dynamic pressure indicative of stronger expansion in the inner heliosphere, and a magnetic-field profile shallower than Parker-spiral expectations. From our catalog ≈55% of ICME events were associated with a preceding shock and sheath, while 45% were not shock-associated. Relative to the ambient solar wind, ICMEs with sheaths are, on average, ≈27% longer in duration, ≈68% more magnetized, ≈30% faster, ≈2.6 times hotter, ≈34% denser, and ≈2.8 times over-pressured than those without sheaths. They also expand ≈96% faster, while maintaining low plasma $\beta$. Additionally, expansion speed strongly scales with ICME size and speed for sheath events, suggesting that sheath dynamics sustain over-pressured ejecta and regulate expansion. For the 32 ICMEs with Heavy-Ion Sensor (HIS) data, ≈30 % satisfy $O^{7+}/O^{6+} \approx 1$. In contrast, Fe indicators ($\langle Q_{Fe} \rangle$, $Fe^{\geq 16^+}/Fe$, $Fe/O$) account for ≈50–65 % of the ICMEs, reflecting solar cycle effects, given that charge-state ratios are generally lower during solar minimum. The superposed epoch analysis shows enhanced heavy-ion signatures within magnetic-cloud intervals, consistent with increased heating and ionization in the low corona. Our catalog therefore provides a high-fidelity, substructure-resolved ICMEs and a physically grounded framework for connecting in-situ composition and dynamics to CME source conditions. This enables multi-mission catalogs, machine-learning applications, data-driven CME evolution models, and space-weather analysis.

## 1. Introduction

Coronal mass ejections (CMEs) are macroscale explosive bursts of plasma, magnetic flux, and helicity that originate from the Sun's atmosphere in a "continuous outflow" of solar wind into the inner heliosphere. Their interplanetary manifestations, as observed in situ, are commonly referred to as Interplanetary Coronal Mass Ejections (ICMEs). ICMEs exhibit diverse in situ signatures as they propagate away from the Sun throughout the heliosphere (Wu & R. P. Lepping 2011; Zurbuchen & Richardson 2006). ICME geoeffectiveness is primarily controlled by the magnitude

[1] Corresponding author: felix.minta@utsa.edu
[2] Department of Physics and Astronomy, University of Texas at San Antonio, San Antonio, TX, 78249, USA
[3] Space Science & Engineering, Southwest Research Institute, San Antonio, TX, 78238, USA
[4] Department of Climate and Space Sciences & Engineering, University of Michigan, Ann Arbor, MI 48109, USA

and orientation of its internal magnetic field (Bothmer & Schwenn 1998; Huttunen et al. 2005); therefore, characterizing ICME evolution and interactions with the solar wind and other structures is essential for understanding the impact of space weather events and improving forecasting and modelling (Hassanin et al. 2024; Kilpua et al. 2017a; Minta et al. 2023).

The term ICME is a generic label that comprises magnetic clouds (MC), magnetic obstacles, and ejecta. Conventionally, MCs form a subset of ICMEs characterized by anomalously low proton temperature and low plasma $\beta$, enhanced magnetic field magnitude, and a smooth, large-scale rotation of the magnetic field, typically lasting ~ a day at 1 au (Gosling et al. 1973; Klein & Burlaga 1982; Richardson & Cane 1995). The magnetic obstacle or ejecta type of ICMEs exhibit complicated structures with disordered magnetic fields and relatively higher plasma $\beta$ or temperature, and are typically faster (>600 km $s^{-1}$) (Burlaga et al. 2001, 2002). They can be associated with multiple interacting CMEs or high-speed streams from several solar source regions (see e.g., Burlaga et al. 2002; Lugaz et al. 2005, 2016; Lugaz and Farrugia 2014) .In such cases, the coronal "memory" of these subsets is not unique and requires careful, independent assessment of the thermodynamic and compositional properties of the plasma. These structural differences, combined with line-of-sight and trajectory effects, complicate attempts to impose a single, universal ICME identification criterion.

About 30% of the fast CMEs observed by coronagraphs have a three-part structure—bright leading edge, a trailing dark cavity, and an internal bright core (Hundhausen 1993; Illing & Hundhausen 1985). These classical parts manifest into a forward shock, compressed turbulent sheath, and the cold, magnetically dominated MC or ejecta. Usually, the shock-heated sheath is characterized by elevated proton temperature, higher plasma $\beta$, enhanced magnetic field magnitude, and strong magnetic fluctuations. Although sheaths are relatively less studied when compared to the shocks and ICMEs, they contribute to several space weather activities, including geomagnetic disturbances (see Liu et al., 2020), compression of the day-side magnetosphere (Kilpua et al. 2019; Lugaz et al. 2016b), long-term depletion of the outer radiation belt electrons (e.g., Hietala et al., 2014; Turner et al., 2019), and acceleration of solar energetic particles (SEPs; Manchester Iv et al., 2005; Lugaz et al., 2016). Note that not all ICMEs exhibit all these substructures. In this paper, we focused on identifying three key substructures: shocks, sheaths, and the MCs.

Over the years, ICMEs have been identified using diverse combinations of single- and multi-signature criteria and visual inspection across spacecraft and heliocentric distances. Early Helios catalogs (0.3–1 au) used magnetic field enhancements, depressed proton temperatures, and Forbush decreases in energetic particles (>60 MeV protons) (Bothmer & Schwenn 1998; Cane et al. 1997), while Richardson and Cane (2010), and Jian et al. (2018) applied similar multi-signature approaches at 1 au using ACE, Wind, and STEREO data. Beyond 1 au, Ulysses and Juno catalogs relied largely on visual inspection of magnetic and plasma properties (Davies et al. 2021; Du et al. 2010; Ebert et al. 2009). In contrast, inner-heliosphere and planetary missions such as MESSENGER and Venus Express identified ICMEs primarily from magnetic field signatures due to instrumental limitations (Davies et al. 2021; Good & Forsyth 2016a). This practice of bypassing plasma signatures may bias catalogs toward MC events. Thus, most early ICME catalogs were therefore expert-driven and visually compiled, introducing subjectivity and inter-observer variability. For example, Richardson (2014) found that only 102 of 279 Ulysses ICMEs overlapped with the catalogs of Ebert et al. (2009) and Du et al. (2010), highlighting significant discrepancies despite using the same dataset. Such inconsistencies, together with the time-intensive nature of manual classification, may delay catalog updates and hinder the application of machine learning models that depend on reliable labeled datasets.

In this study, we implement an automatic ICME detection algorithm that leverages a sliding time window technique with multi-criteria thresholds utilizing high-resolution plasma and magnetic field data from the Proton and Alpha Sensor (PAS; Owen et al. 2020) and MAG (Horbury et al. 2020) instrument, respectively, to identify ICME substructures observed by Solar Orbiter. Thus, we present a comprehensive, continuously updatable ICME catalog from Solar Orbiter which orbits between 0.3 and 1 au with minimal subjectivity and validate it using statistical analysis. Section 2 describes radial variation of the physical parameters used in the algorithm. Section 3 presents the automated identification framework and its implementation for detecting ICME substructures (shock, sheath, and MC). Also in this section, we briefly describe three categories of events identified by the algorithm. Section 4 presents the statistical analysis and discussion of the ICME plasma and composition properties in our catalog. In Section 5, we present the summary, conclusions, and caveats for interpreting the findings.

## 2. Radial Dependence of Physical Quantities

We obtained the MAG Level 2 (L2) magnetic-field vectors in the Radial–Tangential–Normal (RTN) coordinate system ($B_R$, $B_T$, $B_N$), while proton density ($n_p$), temperature ($T_p$), and velocity components ($V_R$ , $V_T$, $V_N$) were extracted from the L2 ground-calibrated moments from PAS. The analysis uses data spanning January 2020 to February 2025. For all vector quantities (magnetic field and velocity), we calculated their vector magnitudes ($|B|$, $|V_p|$). When available, we also retrieved heavy-ion composition and charge-state parameters from Heavy Ion Sensor (HIS) for ICME intervals identified during its operational window (January 2022–April 2023). The composition and charge-state measurements were used for post-detection validation of the ICME catalog.

To implement a robust distance-invariant ICME detection algorithm, we normalize the plasma and magnetic-field parameters to 1 au, effectively removing the background radial trend. This is because Solar Orbiter samples the solar wind over a wide range of distances (≈0.3–1 au), where these parameters vary significantly with distance. Figure 1 presents the radial distributions of $T_p$ (Figure 1(a)), $n_p$ (Figure 1(c)), and dynamic pressure $P_{dyn}$ (Figure 1(d)) for the period under consideration. To quantify the radial trends, we fit each parameter with a power-law of the form $X(r) = a\, r^{\alpha}$ using nonlinear least-squares optimization. The fitting is performed on 0.02-au binned averages in radial distance to reduce the influence of uneven sampling and short-timescale fluctuations. Then, the derived scaling was used to de-trend the data by removing the radial dependence, allowing all measurements to be compared on an equal footing independent of spacecraft distance. For the parameters considered in this study, we obtain radial scalings for $n_p \sim r^{-2.047}$, $|B| \sim r^{-1.521}$ (not shown here), $T \sim r^{-0.569}$, $P_{dyn} \sim r^{-1.890}$, while the plasma $\beta$ shows a positive trend with $\beta \sim r^{0.551}$ (not shown here). In Figure 1(b), the normalized temperature ($T_{norn}$) is subsequently used to derive an empirical relationship with solar wind speed, allowing the reconstruction of the expected temperature ($T_{ex}$) as a function of both heliocentric distance and bulk speed.

Figure 1(a) shows proton temperature decreases with heliocentric distance more slowly than expected for adiabatic expansion, indicating sub-adiabatic cooling during the rising phase of SC25. This behavior departs from earlier studies (e.g Stansby et al. 2018; Perrone et al. 2019, 2022; Liu et al. 2024), suggesting that wave heating, turbulent dissipation, and collisionless transport continue to modify proton population in the inner heliosphere. Figure 1(c) shows that proton density closely follows the expected decrease for a spherically and radially expanding plasma, $n_p \propto r^{-2}$. This agrees with Helios measurements (e.g., Stansby et al. 2018; Perrone et al. 2019), though it is steeper than the slightly flatter trends ($r^{-1.8}$) reported by Perrone et al. (2022) using observations from Solar Orbiter and Parker Solar Probe. In Figure 1(d), we found that dynamic pressure decreases steeply with heliocentric distance, indicating that ICME expansion is strongest in the inner heliosphere, where the external confinement drops most rapidly. This is consistent with previous ICME studies showing that internal ICME overpressure drives radial expansion more rapidly closer to the Sun (see Steiger and Richardson 2006).

For magnetic field (not shown here), the radial dependance is quite shallower than the Parker spiral expectation ($r^{-2}$), likely due to plasma compression from deceleration, Alfvénic fluctuations and turbulence during expansion in the inner heliosphere (e.g., Elliott et al. 2012; Bruno and Carbone 2013; Matteini et al. 2015). The increasing plasma $\beta$ trend occurs because magnetic pressure falls faster than thermal pressure, making the ambient wind less magnetically dominated outward (see e.g., Perrone et al. 2022). Overall, our gradients are steeper because Solar Orbiter samples solar wind closer to the Sun where radial evolution is more rapid and compression or rarefaction diverges (see e.g., (Elliott et al. 2012). Also, the discrepancies between our scaling and results from earlier missions (e.g., Helios, Ulysses, ACE) in the previous studies are attributed to variations in solar wind conditions, dynamic interaction regions between fast and slow solar wind streams, and the influence of secondary proton populations. Nevertheless, this empirical approach ensures: (i) the removal of background radial trend; (ii) the use of uniform detection thresholds across the inner heliosphere; (iii) consistent application of ICME identification criteria developed for 1 au observations, while enabling direct comparison with established 1-au ICME catalogs; (iv) clearer separation of ICME signatures from the ambient solar wind. Therefore, statistically characterizing the expected state of the solar wind, any significant deviation in the normalized parameters can be more reliably attributed to transient structures such as stream interaction regions (SIRs), rather than to the spacecraft location or to the ICMEs themselves.

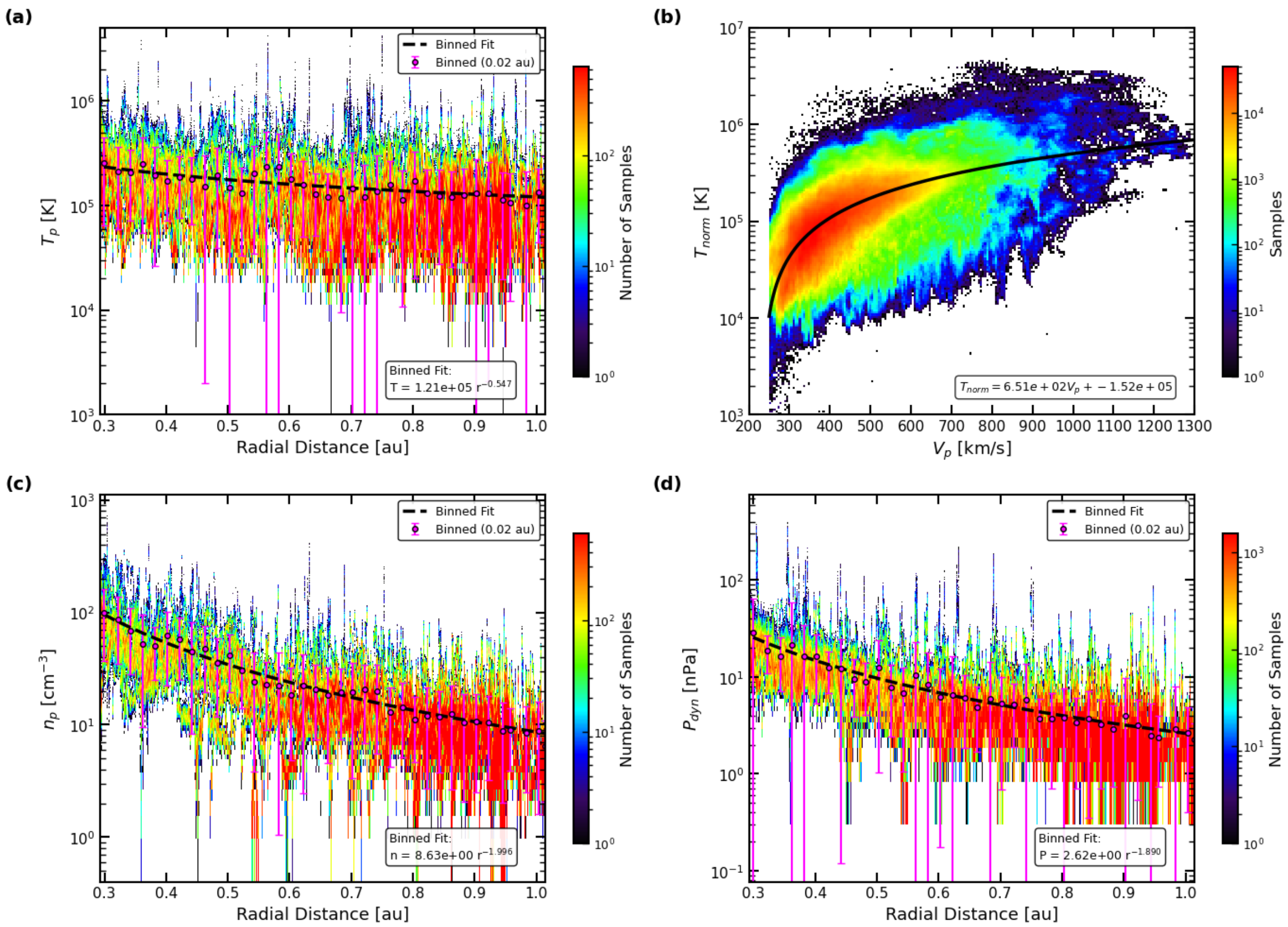

Figure 1. Radial profiles illustrating the expected decrease in $T_p$, $n_p$, and $P_{dyn}$ with increasing heliocentric distance. **Top left**: proton temperature $T_p$ as a function of heliocentric distance, with measurements (colored by sample count), 0.02-au bin averages (magenta), and power-law fits to the binned values (black). **Top right**: proton bulk speed $V_p$ and normalized temperature $T_{norm}$, with empirical fit $T_{exp,norm}$ overlaid. **Bottom left**: proton number density $n_p$ versus radial distance. **Bottom right**: dynamic pressure $P_{dyn}$ versus radial distance.

## 3. Algorithm Architecture

In practice, the automated ICME substructure detection algorithm (summarized in Figure 2) reads in PAS and MAG data. To ensure consistency across data with varying temporal resolutions, the parameters are resampled to a uniform 10-minute cadence using linear interpolation. Next, the algorithm utilizes a sliding window technique to systematically scan the time series data in overlapping 36–hour window in 1–hour steps and delineate areas where initially the total magnetic field ≥ 7nT, consistent with ICME magnetic-enhancement thresholds used in previous studies (Huttunen et al. 2005; Kilpua et al. 2012; Lepping et al. 2005; Rüdisser et al. 2025) and the plasma $\beta$ is low (*i.e.,* $< 0.8$) (see Burlaga et al. 1981). These conservative pre-filters efficiently isolate regions of magnetically enhanced, low plasma $\beta$, typical of ICME candidates, while remaining sensitive to ambient solar wind conditions. Continuous stretches of data that meet the initial conditions are subject to duration checks. Thus, intervals between 4 and 72 hours (e.g., Huttunen et al. 2005; Good and Forsyth 2016; Davies et al. 2021) are retained and further evaluated for substructure signatures using multi-signature criteria. The algorithm next assesses the upstream conditions over a 12-hour period preceding each candidate interval to determine relative enhancements and variances in magnetic field magnitude across the downstream region. All intervals passing these pre-filters are simultaneously evaluated for both non-MC ICME signatures and MC flux-rope signatures.

### 3.1 Detection of Non-MC ICME Events

Non-MC ICMEs are identified using a multi-criteria approach that does not require coherent large-scale field rotation. A candidate interval is classified as a non-MC ICME when at least three of the following six well-established signatures are simultaneously present: (i) enhanced magnetic field (Klein & Burlaga 1982), (ii) anomalously low proton temperature ($T_{norm}/T_{ex} < 0.5$) (Gosling et al. 1973; Richardson & Cane 1995), (iii) low normalized plasma $\beta$ ($\beta_{norm} \leq 0.1$) (Klein & Burlaga 1982; Lepping et al. 1990), (iv) declining speed profile (Klein & Burlaga 1982; Russell & Shinde 2003; Zurbuchen & Richardson 2006), (v) significant change in magnetic field direction ($\theta_{RTN} \geq 45°$) (Klein & Burlaga 1982; Lepping et al. 2005), and (vi) low magnetic field variance (Klein & Burlaga 1982) relative to the upstream solar wind. Some of these non-MC ICMEs are 'cloud-like', showing evidence of coherent magnetic field rotation without fully meeting the strict criteria of an MC applied here (Burlaga et al., (1981); Klein and Burlaga, (1982). A similar observation was also made by Cane and Richardson, (2003). Because no single signature is universal or sufficient (Ebert et al. 2009; Gosling 1990; Zurbuchen & Richardson 2006), our algorithm combines multiple downstream indicators with flux-rope signatures rather than relying solely on plasma conditions. This approach ensures that the algorithm automatically ignore magnetically enhanced CIRs and SIRs, which lack the characteristic flux-rope magnetic configuration (see e.g., Good and Forsyth, 2016).

### *3.2 Detection of MC Events*

Magnetic clouds are identified within the same set of candidate intervals when a smoothly rotating, enhanced magnetic field characteristic of a flux-rope structure is present. The algorithm detects these in two complementary steps. First, the algorithm rigorously assesses whether the downstream region shows smooth, and large-scale enhancement in the magnetic field. This is achieved through a comprehensive component-wise regression analysis. For each component within a candidate interval, the algorithm quantifies the smoothness, and enhancement in the magnetic fields using the coefficient of determination ($R^2$) and the total change in field strength ($\Delta B$), respectively. Specifically, the primary method uses a quadratic fit to better model the often non-linear, curved trajectory of the magnetic field vector in the flux rope. This is particularly effective for capturing the "S-shaped" hodograms characteristic of MCs. Linear regression remains an option for longer-duration and more uniform rotations, since such MCs often have reduced magnetic-field curvature and become more linear. Note that the quality of the fit is a central parameter for establishing the smoothness required for MC identification.

Second, the algorithm directly examines magnetic field rotation by calculating the polar $(\theta_{RTN})$ and azimuthal $(\phi_{RTN})$ angles. It then assesses their smoothness and monotonicity over several hours to confirm the presence of a coherent flux-rope-like orientation. This provides a good measure of directional continuity, ensuring robust and physically meaningful MC identification. We classify MCs into two categories based on the strength and clarity of their flux-rope signatures. "Good-MC" events show typical flux-rope characteristics: strong field enhancement, depressed plasma parameters, and clear smooth rotation. In contrast, "Complex flux rope" events show disordered magnetic rotation, often accompanied by weaker field strength and/or elevated proton temperature and plasma $\beta$ (see e.g., Burlaga et al. 2001, 2002; Lugaz et al. 2005, 2016; Lugaz and Farrugia 2014). In practice, a "Good-MC" is identified when the criteria satisfy either $R^2 \geq 0.5$ or a change in $\theta_{RTN} \geq 45°$ with $\Delta B > 7$nT, and either $\beta_{norm} < 0.1$ and/or $(T_{norm}/T_{ex}) < 0.5$ during the candidate interval. These thresholds were iteratively optimized to maximize sensitivity to flux ropes, while maintaining high specificity against non-ICME structures. However, some events show complex structures. For example, downstream regions may display organized magnetic rotation but lack clear plasma signatures (i.e., no significant decrease in $\beta_{norm}$or $T_{norm}$), while other cases may show low plasma $\beta$ or temperature but highly disordered magnetic field orientations. Such events are categorized as "Complex flux rope". To account for such cases, we adopt moderately relaxed criteria in the algorithm, placing greater emphasis on the robustness of the smooth rotation signature when other parameters are less definitive. Thus, "Complex flux rope" types are required to have (i) $R^2 \geq 0.3$ and $\theta_{RTN} \geq 55°$ and (ii) [$\beta_{norm} < 0.13$ or (iii) $(T_{norm}/T_{ex}) < 0.5$ or $|B| > 8$ nT] and (iv) $\Delta B > 5$ nT.

### *3.3 Practical Tuning and Caveats*

During the algorithm's development, we observed that the sliding-window approach frequently generates multiple overlapping candidate intervals for the same physical event. To resolve these and produce a clean catalog, the algorithm employs a hierarchical merging logic that consolidates overlapping intervals of the same type. For adjacent intervals from different candidate types, the algorithm merges and classifies the event based on the highest-quality signature present within that interval. For example, if a "Good-MC" and a "Complex-flux rope " overlap, the entire interval is classified as a "Good-MC". This ensures that the most unambiguous signatures dictate the final event characterization and prevent the dilution of high-fidelity detections.

To this end, all identified non-MC ICMEs, and MCs are constrained to durations of 4 and 72 hours. Although a 24-hour minimum is typically expected for MCs at 1 au (Burlaga et al. 1981b), this threshold may overlook clouds encountered closer to the Sun, where their spatial and crossing times are generally shorter. Because Solar Orbiter frequently operates much nearer to the Sun, we adopt a minimum duration of 4 hours to capture large-scale flux ropes likely associated with CMEs in our catalog, while discarding smaller, non-CME-related flux ropes commonly present in the solar wind (see Moldwin et al. 2000; Janvier et al. 2014). It is important to note that ICME/MC duration remains subjective as it depends on several factors, including (i) the intrinsic size and orientation of the event, (ii) the spacecraft trajectory (e.g., central axis vs. the flank crossing); and (iii) radial evolution from the Sun during ICME expansion. For instance, a typical ICME at 1 au could have a maximum duration ≈36 hours; however, Steiger and Richardson, (2006) stated that an overly pressured ICME could expand to 5.2 days at larger heliocentric distances (~5.3 au), particularly during flank encounters (Lugaz et al. 2020b; Marubashi & Lepping 2007). Based on these considerations, we adopt a conservative maximum ICME/MC duration of 72 hours to retain the largest and most slowly evolving events while minimizing contamination by non-ICME structures. Lastly, the boundaries of our detected events are consistent with methods described by Kilpua et al., (2012), which mainly rely on magnetic field rotation signatures (except in the presence of a shock) rather than temperature and plasma $\beta$. This choice is motivated by the fact that short intervals of low temperature and plasma $\beta$ potentially occur outside ICMEs (see Richardson and Cane 1995; Lavraud et al. 2010).

### *3.4 Identification of Shocks*

In this subsection, our algorithm follows a similar approach outlined in Appendix A of Trotta et al. (2025), with slight modifications to look for forward shocks by near-simultaneous abrupt jumps in magnetic and plasma parameters ahead of the ICMEs. The detection process relies on a sliding window approach, where the algorithm steps through the data in increments of 2 minutes, analyzing upstream and downstream regions at each time step. For each time point, the algorithm defines upstream and downstream regions separated by a 30-second exclusion zone to avoid contamination from the shock transition itself. A 3-minute averaging window is used to compute mean values of $|B|$, $|V_p|$, *and* $n_p$ in these regions, as this duration is sufficiently long to provide stable averages while remaining short enough to resolve shock-related discontinuities.

The core of the shock identification logic relies on evaluating jump conditions across the shock front. For the magnetic field, a compression ratio of at least 1.2 (i.e., $|B|_{down}/|B|_{up} \geq 1.2$) (Kilpua et al. 2015; Neugebauer et al. 2003; Trotta et al. 2025) is required, consistent with the physical expectation of magnetic field compression in fast forward shocks. When plasma data are available within an 8-hour window around the analysis time, the algorithm further checks for jumps in plasma density (i.e., $n_{p,down}/n_{p,up} \geq 1.2$) and a velocity difference of at least 20 km s$^{-1}$ between downstream and upstream regions (i.e., $V_{p,down} - V_{p,up} \geq 20$ km s$^{-1}$) (Kilpua et al. 2015; Neugebauer et al. 2003; Trotta et al. 2025). The use of a velocity difference, rather than a strict compression ratio, accounts for the complex variability of solar wind flows, where velocity changes may not scale proportionally, but remain indicative of shock formation. If plasma data are unavailable within the 8-hour window, the algorithm fall back to using only the magnetic field jump condition. This fallback mechanism is critical for robustness, as plasma data from PAS contains gaps. In such cases, detections are retained but interpreted with lower confidence, as plasma parameters provide important complementary diagnostics. A shock is therefore confirmed when all available jump conditions are satisfied, or when the magnetic field criterion alone is met in the absence of plasma data.

### *3.5 Sheath Confirmation*

The sheath region is the compressed and turbulent plasma between an interplanetary shock and the leading edge of an ICME, often characterized by high plasma $\beta$ and significant magnetic field fluctuations. The algorithm identifies sheaths as follows: For ICMEs preceded by multiple shocks, the algorithm first ranks the shocks to identify the one with the highest compression ratios as the associated driving shock of the non-MC ICME/MC. Subsequently, the region between the selected shock's time and the leading edge (i.e., non-MC IMCE/MC start time) is defined as the sheath region. For each sheath identified, the algorithm evaluates the following conditions for confirmation: (1) the average plasma $\beta$ must be greater than unity; (2) average values of density and temperature within the sheath must be significantly greater than the averages within the ICME/MC; (3) higher variance of plasma $\beta$ and density within the sheath region, reflecting its turbulent and compressed nature; (4) variance of the total magnetic field strength, to quantify the degree of magnetic fluctuations. The sheath thickness and duration depend on several factors, including the shock compression ratio, velocity, and physical characteristics of the driving ICME (e.g., shape and curvature radius; see e.g. Mulligan and Russell, 2001), and may even expand from the ICME nose toward the flanks (Kilpua et al. 2017b). Therefore, no strict physical and time constraints are imposed on sheath regions in our algorithm.

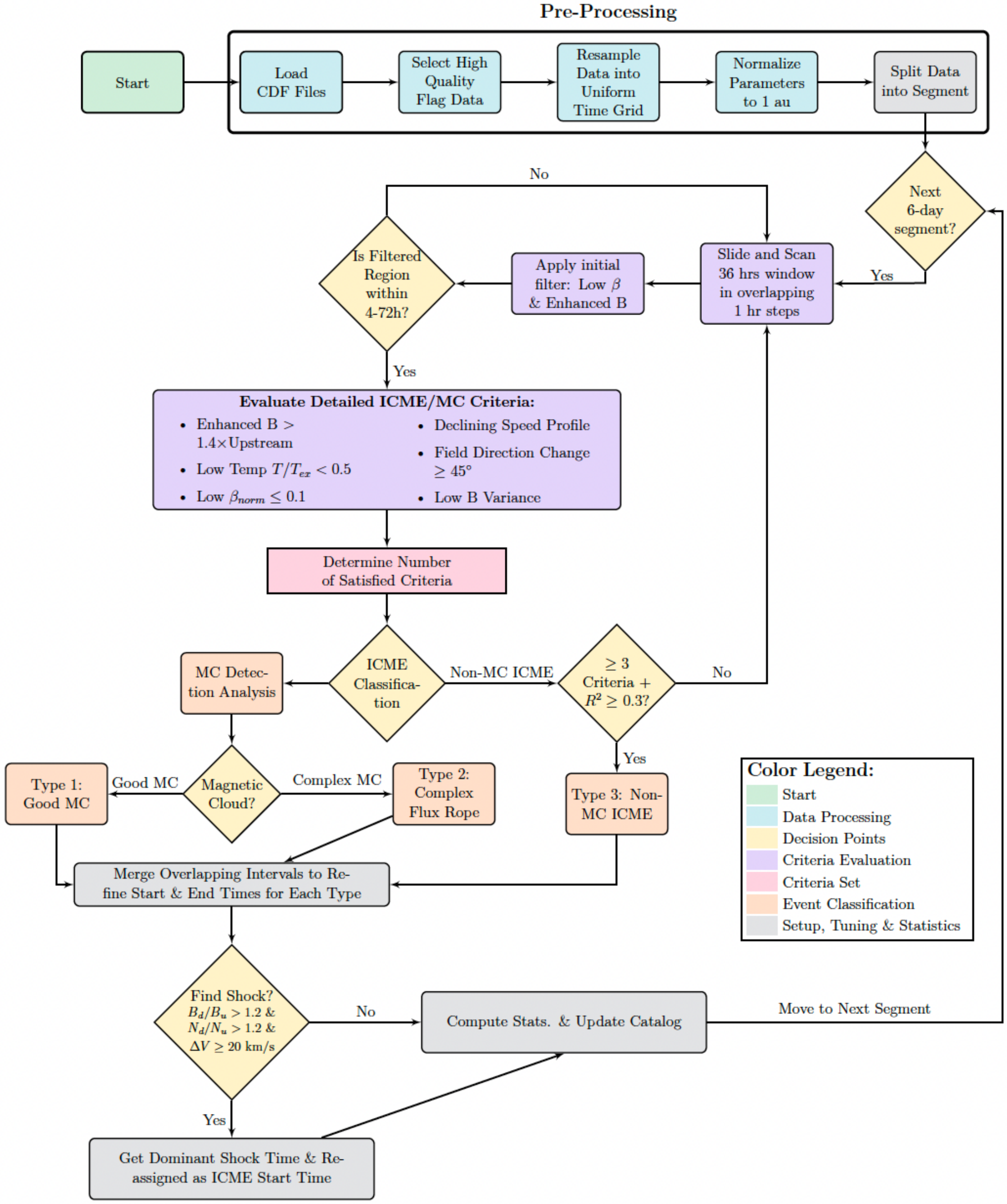

Figure 2: Shows the flowchart of the automated ICME substructure detection process, which involves data preprocessing, sliding-window candidate selection, ICME/MC classification, interval merging, and subsequent shock and compilation of the catalog.

## *3.6 Overview of ICME Event in each category*

### *3.6.1 Good–MC Event*

Figure 3 presents an example of a Good–MC ICME event observed by Solar Orbiter between 2024 June 19 10:16 UT and June 21 07:57 UT. The figure is organized into panels (a–e), showing the magnetic field, azimuthal angle, plasma speed and density, temperature, and plasma beta, respectively. The ICME substructures identified by our algorithm are highlighted, with the shock marked by a red dashed vertical line, followed by the sheath (gray) and the MC (cyan). This event satisfies all established ICME identification criteria summarized in Figure 2. For comparison, the ICME intervals reported by ICMECAT (black; Möstl et al. 2017, 2020) and LASSOS (purple; Nieves-Chinchilla et al. 2016) horizontal bars are overplotted above panel (a). In this case, the MC interval agrees closely with both external catalogs, although ICMECAT does not include the shock and sheath, and the LASSOS interval ends about 5 hours earlier than our ICME end time.

As shown in Figure 3, our algorithm identifies a clear shock ahead of the leading edge (panel a), characterized by an abrupt increase in the magnetic field magnitude ($|B|$), accompanied by simultaneous jumps in bulk speed and proton density (panel c). In the downstream region of the shock, the sheath (lasting ~13 hrs.) exhibits strong fluctuations in both magnetic and plasma parameters, consistent with a compressed and turbulent solar wind environment. However, these quantities decline more gradually through the ICME, indicating the slow self-similar expansion on the order of 25 km s$^{-1}$. The MC shows a relatively smooth, though modest, rotation in the magnetic field direction (see panel b), suggesting a weakly organized flux rope structure. The mean magnetic field strength within the MC is ≈12.4 nT, with a peak value of ≈24.5 nT, further emphasize a moderately enhanced magnetic structure. In panel (d), $T_p$ (black curve) shows a pronounced depression relative to $T_{ex}$ (red curve), indicating non-adiabatic cooling within the MC. This is consistent with the low plasma $\beta$ values observed in panel (e), where $\beta$ remains significantly below unity for most of the MC interval.

The discrepancy in ICME start times (omissions of shock and sheath) in ICMECAT reflects that ICMECAT relies solely on organized magnetic field orientation: i.e., (i) enhanced total magnetic field and (ii) rotation in the field components (Möstl et al. 2017, 2020). In contrast, our algorithm uses at least three independent plasma and magnetic signatures, yielding a more complete and physically consistent identification of ICME substructures. This broader multi-step approach is more robust and inclusive, especially in cases where the magnetic field signatures alone may be ambiguous due to observational geometry or solar wind interactions. Finally, it is worth noting that event occurred after the HIS failure (April 2023), and therefore heavy-ion composition data are unavailable. Despite this limitation, the combination of magnetic and plasma signatures provides a confident classification of this event as a Good–MC ICME.

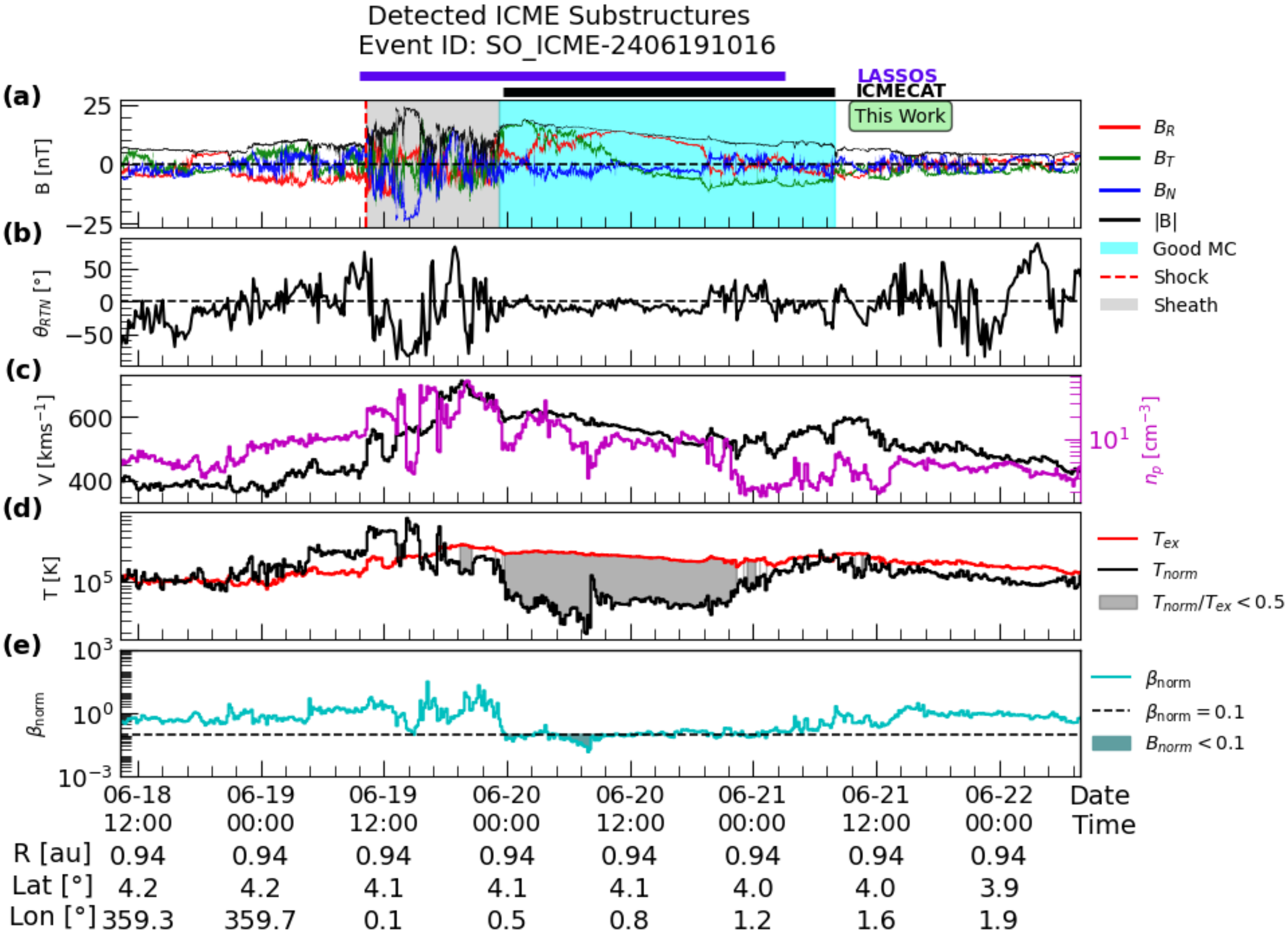

Figure 3. Overview of a Good-MC observed by Solar Orbiter from 2024 June 19 10:16 - 2024 June 21 07:57 UT at 0.94 au. The ICME substructures include the sheath (gray) bounded by a forward shock (red dashed line) and the MC-like ejecta (cyan). Panels (a-e): the total magnetic field and RTN components (This work), the magnetic field polar angle, proton speed (black) and density (magenta), proton temperature (black) with expected solar wind temperature (red; ×0.5 shown, gray), and plasma $\beta$ (cyan). Black and Purple bars above panel show the ICME intervals listed in ICMECAT and LASSOS catalogs, respectively.

### *3.6.2 Complex Flux Rope Event*

Figure 4 shows an ICME tagged as a complex flux rope, for which our algorithm identified the shock, sheath, and flux rope-like structures. Similar to Figure 3, panels (a–e) show the magnetic and plasma properties, while the horizontal bars above panel (a) indicate the ICME boundaries reported by ICMECAT and LASSOS for this event. In addition, panels (f–j) extend the analysis to include heavy-ion composition properties. In contrast to the well-organized MC discussed earlier in Figure 3, this event exhibits a more intricate structure in which magnetic and plasma signatures do not satisfies the strict criteria for “Good-MCs”.

The shock arrives at Solar Orbiter on 2022 June 17 00:40 UT, followed by a ≈ 16 hr. sheath region and a coherent magnetic structure between 2022 June 17 16:10 UT and 2022 June 19 00:19 UT. The sheath is extended and dynamically active and appears to encroach into the leading edge of the flux rope. Within the complex flux rope, the magnetic field strength remained elevated ($|B| \approx$ 15 nT, with a maximum of ≈22 nT) and shows a clear orientation at the center but somewhat irregular rotation at the trailing edge (panel a and b). However, the plasma behavior departs from the typical signatures of an MC. Thus, the downstream region shows higher $T_{norm}$ (panel d), density (panel c) and moderately elevated plasma $\beta$ (panel e) relative to the upstream solar wind. In panel (c), bulk speed profile indicates that the flux rope is undergoing irregular expansion at an average expansion speed of ≈ 36 km $s^{-1}$ and a mean speed of ≈ 406 km $s^{-1}$. These features suggest that flux rope is not fully detached from the ambient solar wind. Instead, the leading edge appears to be affected by a mixing layer, consistent with magnetic reconnection properties (i.e., high temperature, density, and plasma $\beta$). Thus, the flux rope's geometry agrees with previous studies, where an increase in these three high-state plasma signatures occasionally coincides with organized large-scale magnetic rotation and rear-edge cloud erosion at 1 au (see Ruffenach et al. 2012, 2015).

From Figure 4, ICME start time agrees with ICMECAT and LASSOS, though both catalogs reported the end time ≈32 hours earlier than ours, excluding a substantial portion of the trailing edge (see bars above panel a). Therefore, we include heavy ion composition data to support our boundary detection. In panels (f and g), carbon and oxygen states, respectively, remain elevated and steady throughout the ICME before returning to ambient levels. Similarly, *Fe* charge states (panel h) are enhanced but with fluctuations near the trailing edge. In panel (i), $O^{7+}/O^{6+}$ ratio (black) and *Fe/O* abundance ratio (blue) also remain elevated on average compared to the ambient. While $C^{6+}/C^{5+}$ ratio (black) shows significant enhancement and varies across the ICME interval, $C^{6+}/C^{4+}$ ratio increases only near the trailing edge (see panel j). Based on these observations, we agree that our automatic algorithm made a fair judgment to terminate the ICME at the time composition signatures return to normal ambient solar wind conditions.

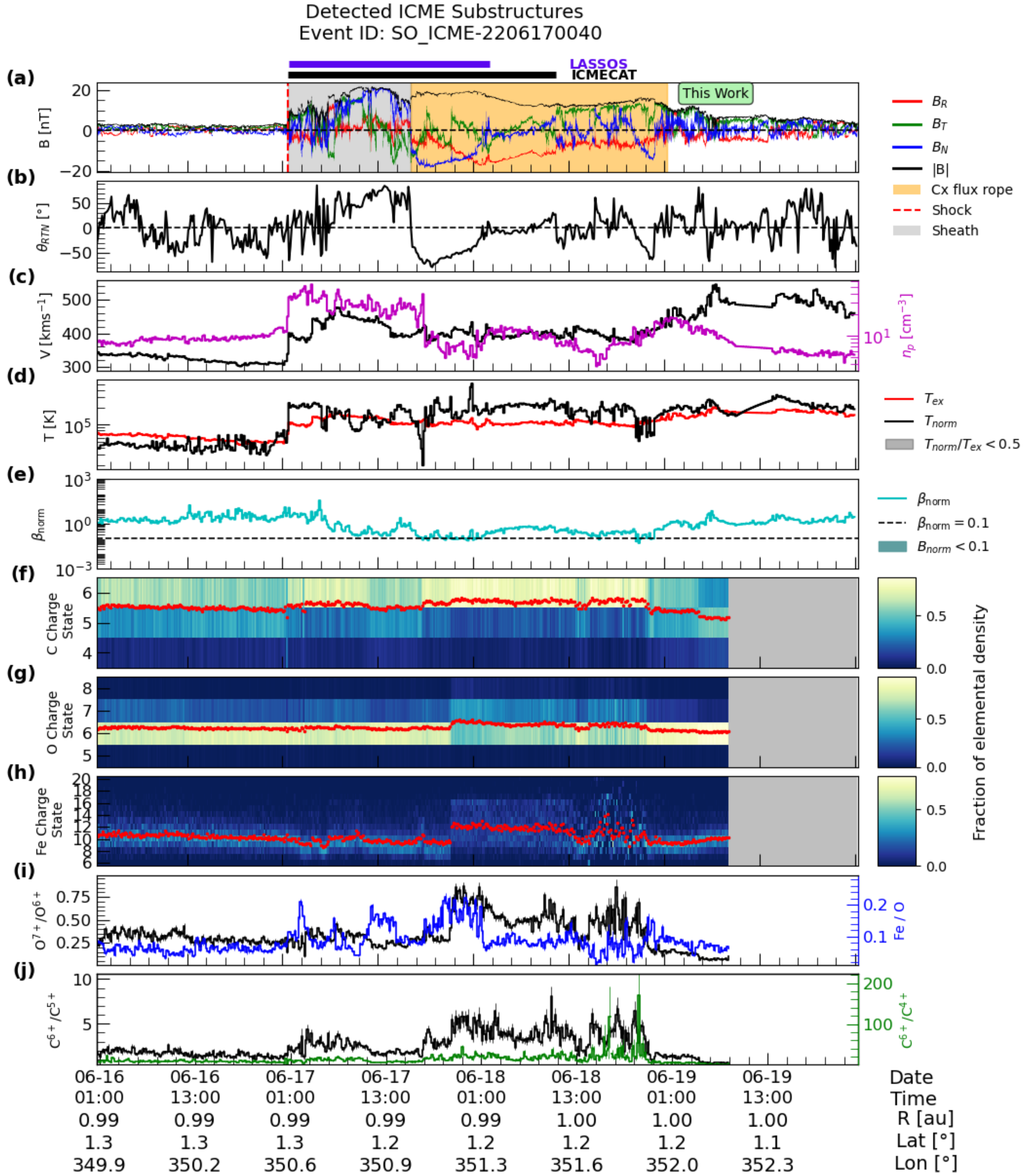

Figure 4. Complex flux rope ICME event observed by Solar Orbiter at ≈ 1au (2022 June 17- 2022 June 19). Panels (a-e) are the same as Figure 3, except that the ICME /ME identified by the algorithm in this case is marked as "Complex flux rope". Panels (f-h) show carbon, oxygen, and Fe charge states (red lines are averages), respectively. Panel (j) shows $O^{7+}/O^{6+}$ (black) and *Fe/O* ratio (blue) and Panel (j): $C^{6+}/C^{5+}$(black) and $C^{6+}/C^{4+}$(green) ratios.

### 3.6.3 Non-MC ICME Event

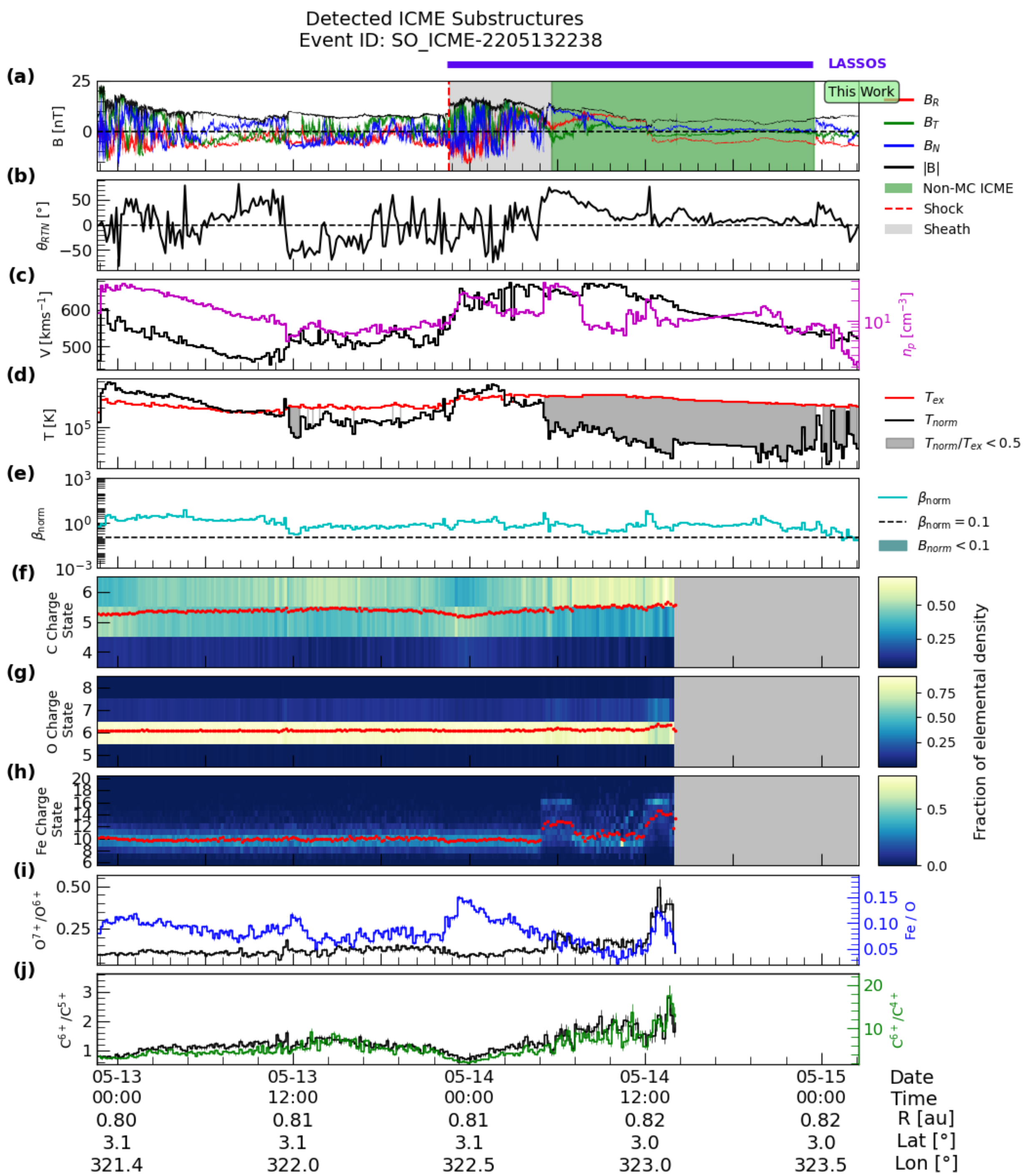

Figure 5. Non-MC ICME events observed by Solar Orbiter from 2022 May 13 22:38 UT and 2022 May 14 23:31 UT. See Figure 4 for descriptions of each panel.

In Figure 5, we present an example of an ICME detected and tagged as a non-MC ICME (panel a, green shading). Note that the panels follow the same format as shown in Figure 4. There is no corresponding ICME is reported in ICMECAT, however, the event is cross listed in the LASSOS catalog with nearly identical start and end times (above panel a, purple bar). In panel (a), a fast-mode shock is observed at 2022 May 13 22:38 UT, immediately followed by a disturbed sheath region that persists until 2022 May 14 05:036 UT. The ICME interval continues until 2022 May 14 23:31 UT. During this interval, the magnetic field magnitude remains coherent but relatively weak ($|B| \approx 8.4$ nT) compared to the background and does not display a significant smooth rotation (see panel b). This value is significantly higher than the maximum magnetic field magnitude of 5.2 nT reported for the ICME event on 2008 July 25 13:25 –23:30 UT (see Fig. 1; Kilpua et al. 2011). It is worth noting that that the ICME presented here satisfied three sets of criteria– a monotonic decline in speed (panel c) indicative of self-similar expansion at speed of $\approx$ 62 km s$^{-1}$, significant depression in $T_{norm}$ relative to $T_{ex}$ (panel d), and relatively low magnetic field variance compared to the ambient (see panel a). In Figure 5, it can be observed that the upstream region is characterized by slightly reduced temperature of $\approx$ 24 hours in the absence of other classical ICME signatures. However, the downstream region is characterized by weak magnetic field, with reduced level of fluctuations, and mean solar wind speed of $\approx$ 615 km s$^{-1}$. Taken these together, arguably suggest that the coherent ICME structure is embedded within the ambient solar wind (cf. Kilpua et al. 2011). Overall, this example demonstrates the algorithm's ability to identify weak ICMEs with little to no flux rope structure, which might be overlooked through visual inspection, highlighting its performance relative to existing catalogs.

## 4. The ICME Catalog and Comparison with External ICME Lists.

Using our automatic algorithm, we compiled 138 ICMEs between October 2020 and February 2025 observed by Solar Orbiter, summarized in Figures 6 and 7. The full catalog, which includes magnetic-field, plasma, expansion, sheath, and composition parameters, is publicly available in Zenodo[3], and a concise description of all quantities is provided in Table 1 of Appendix A. From the catalog, ≈55% have multi-substructures (i.e., shocks, sheaths, and MCs) while ≈45% are standalone MCs. By way of classification, Figure 6(a) shows that Good-MC/ICMEs comprised the largest group, ≈57%, of which ≈57% possess sheaths, and ≈43% do not. Similarly, Complex-flux ropes accounted for ≈12%, with ≈65% sheath-associated and ≈35% without sheaths. Non-MC ICMEs comprised ≈31%, of which ≈48% contained sheaths, while ≈52% were non-sheath events. Overall, Good and Complex flux ropes represent ≈70%, with the remaining ≈30% classified as non-MC ICMEs. These fractions emphasize findings by Cane and Richardson (2003); Richardson and Cane (2004), who reported that ≈70% of ICMEs during solar minimum and nearly 20% during solar maximum contain MCs.

Figure 6(b) compares yearly ICME frequencies listed in our catalog (blue), ICMECAT (purple), and LASSOS (orange) restricted to Solar Orbiter events over the same interval. Overlayed on the bars are events common to our catalog and ICMECAT (cyan) or LASSOS (lime). Here, ICME events are considered common when the absolute difference in start or end times between catalogs is ≤ 6 hours. During 2020–2025, our catalog contains 138 compared to 140 in ICMECAT and 71 magnetic obstacles in LASSOS, which is substantially smaller, representing roughly half the number reported in the other two catalogs. Of the 140 ICMEs in ICMECAT, ≈64 % overlap with our list, and ≈65 % of our events coincide with ICMECAT entries. This implies ≈46 % of our events are unique, while ≈39 % are unique to ICMECAT. These differences underscore agreement between the catalogs while highlighting the significant subjectivity and methodological differences inherent in ICME identification. Thus, the multi-signature-based algorithm is inherently more inclusive and sensitive to ICME classes that can be missed when selection is based solely on organized magnetic-field rotations. In particular, the algorithm reliably identifies: (1) ICMEs with weak magnetic fields; (2) events whose signatures are partially obscured due to strong interaction with the ambient solar wind; and (3) ICMEs whose basic confirmation comes from heavy-ion composition metrics not used in ICMECAT's selection criteria.

[3] https://doi.org/10.5281/zenodo.20262627

Similarly, of the 71 MOs listed in LASSOS catalog, ≈ 38% ICMEs are associated with our catalog, while nearly ≈ 20% of events in our catalog appear in LASSOS. Thus, ≈80% of our events are not in LASSOS, while 62% of events in LASSOS are not present in our catalog. We note that these disagreements are remarkably significant but can be attributed again to differences in methodologies. For instance, LASSOS catalog was created through a test case application of the reconstruction technique with a generalized circular-cylindrical model for only flux rope structures (see Nieves-Chinchilla et al. 2016). Arguably, these systematic discrepancies highlight the absence of an independent "ground truth" in ICME catalog and emphasize that methodological choices (i.e., signatures, thresholds, and matching rules) strongly shape ICME lists. Nevertheless, yearly ICME counts in all three catalogs also reflect both solar activity and instrument coverage. During solar minimum (~2020), all catalogs show low occurrence, however, the number of ICME events differ (our catalog: 1, ICMECAT: 4, LASSOS: 10), mainly because of limited PAS data coverage included in our algorithm and ongoing calibration early in the Solar Orbiter mission, whereas ICMECAT and LASSOS rely primarily on magnetic field data. During the rising phase of solar cycle (SC) 25, all catalogs show increasing ICME rates tracking the rise in solar activity. Specifically, our catalog and ICMECAT closely mirror each other in 2022 and 2023, indicating strong agreement during periods of moderate-to-high activity.

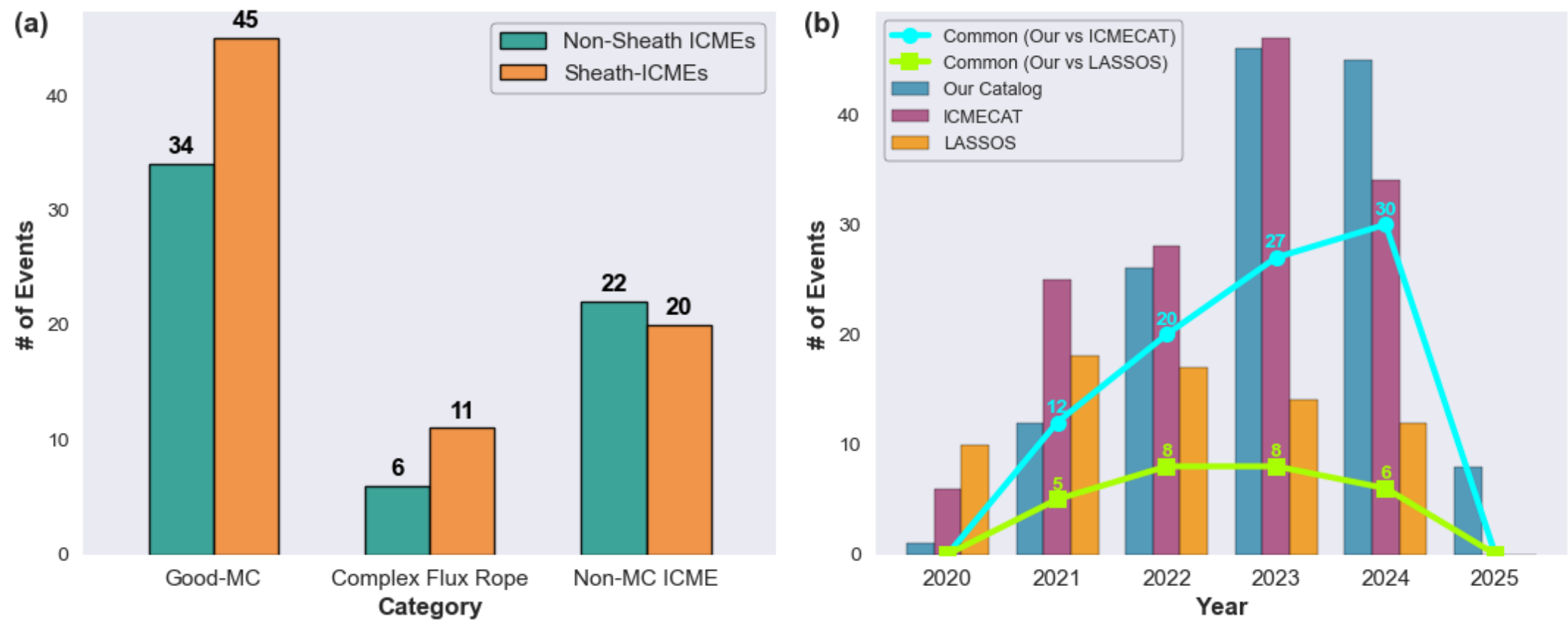

Figure 6. Left: Distribution of ICME events by category and presence of a sheath region. Bars show events without sheaths (teal), and events with sheaths (orange). Right: Yearly comparison and cross validation our catalog (blue) with ICMECAT (purple) and LASSOS (orange). Line curves show common events between catalogs. Note that 2020 and 2025 have only partial plasma data coverage.

**Table 1:** Yearly summary of the ICME catalog. Columns 1–2 list the year and sheath classification (N: without sheath; Y: with sheath). Columns 3–9 give yearly averages of duration, magnetic field ($|B|$), speed ($|V_p|$), temperature ($T$), density ($n_p$), plasma $\beta$, dynamic pressure ($P_{dyn}$), expansion speed ($Vexp$), and Solar Orbiter radial distance ($Rso$). * indicates incomplete ICME data for that year; • indicates that 2020 includes only one event.

| Year | Sheath | ICME ⟨Dur⟩[h] | ⟨B⟩ [nT] | ⟨V<sub>p</sub>⟩ [km s$^{-1}$] | ⟨T⟩ [K] | ⟨n<sub>p</sub>⟩ [cm$^3$] | ⟨$\beta$⟩ | ⟨P<sub>dyn</sub>⟩ [nPa] | ⟨V<sub>exp</sub>⟩ [km s$^{-1}$] | ⟨R<sub>so</sub>⟩ [au] |
|---|---|---|---|---|---|---|---|---|---|---|
| 2020• | N | 13.71 | 7.55 | 405.31 | $4.9 \times 10^4$ | 7.08 | 0.23 | 1.96 | 4.97 | 0.98 |
| 2021 | N | 27.36 | 14.32 | 352.57 | $5.2 \times 10^4$ | 25.02 | 0.24 | 4.88 | 44.66 | 0.72 |
| 2021 | Y | 24.77 | 15.91 | 391.91 | $1.0 \times 10^5$ | 21.7 | 0.46 | 5.78 | 30.23 | 0.79 |
| 2022* | N | 21.82 | 15.1 | 417.38 | $7.9 \times 10^4$ | 24.31 | 0.25 | 5 | 50.98 | 0.76 |
| 2022* | Y | 24 | 28.02 | 521.57 | $2.3 \times 10^5$ | 42.5 | 0.66 | 16.27 | 92.5 | 0.67 |
| 2023 | N | 15.71 | 16.28 | 390.12 | $8.3 \times 10^4$ | 36 | 0.3 | 7.87 | 25.66 | 0.69 |
| 2023* | Y | 23.51 | 26.07 | 503.21 | $2.1 \times 10^5$ | 40.08 | 0.67 | 18.86 | 77.42 | 0.69 |
| 2024 | N | 20.43 | 18.56 | 396.53 | $1.1 \times 10^5$ | 30.08 | 0.26 | 6.63 | 35.5 | 0.67 |
| 2024 | Y | 21.37 | 31.09 | 534.83 | $2.7 \times 10^5$ | 43.19 | 0.69 | 22.83 | 66.46 | 0.65 |
| 2025 | N | 8.77 | 15.33 | 290.55 | $6.9 \times 10^4$ | 40.49 | 0.42 | 5.88 | 21.12 | 0.7 |
| 2025 | Y | 20.16 | 20.96 | 494.4 | $1.5 \times 10^5$ | 30.4 | 0.53 | 10.22 | 25.21 | 0.78 |

Table 1 highlights several results from the yearly summary of 138 events in our catalog. For instance, on average, sheath-associated ICMEs have stronger magnetic fields (15.91–31.09 nT) than non-sheath events (7.55–18.56 nT). These values are consistent with values reported in previous studies (Forsyth et al. 2006; Jian et al. 2018; Kilpua et al. 2017b; Richardson & Cane 2010). Similarly, the average speeds for sheath-associated ICMEs, though slightly higher than those of non-sheath ICMEs, are found to scale with typical solar wind speed (350 – 440 km s$^{-1}$) observed during the minimum and rising phase of the solar cycle, suggesting that CMEs accelerate or decelerate to the ambient solar wind speed (see Kilpua et al. 2017a). Also, non-sheath ICMEs are relatively cooler than ICMEs with sheaths. Thus, temperatures of non-sheath ICMEs generally scale with the ambient solar wind (≈$1.5 \times 10^5$ K), whereas sheath-associated ICMEs show substantially higher values, implying that thermal energy from the shocked sheath plasma is conducted or advected into the core of the ejecta. In terms of density, non-sheath and sheath-related ICMEs show higher than normal typical solar wind density value of 7 cm$^{-3}$ (see Forsyth et al. 2006; Richardson and Cane 2010; Mitsakou and Moussas 2014).

As expected from our selection criteria, plasma $\beta$ generally remains below unity for both groups, consistent with magnetic pressure dominance within ICMEs. Dynamic pressures in non-sheath ICMEs, though lower than ICMEs with sheaths, generally scale with the ambient solar wind dynamic pressure ≈1.4–2.4 nPa (McComas et al. 2013). Expansion speeds in both groups are consistent with values from previous studies (see Wang et al. 2005; Forsyth et al. 2006; Richardson and Cane 2010). Taken together, the yearly averages indicate that sheath-related ICMEs are, on average, ≈27 % longer in duration, ≈68 % more magnetized, ≈30 % faster, ≈2.6 times hotter, ≈34 % denser, and ≈2.8 times more over-pressured than non-sheath ICMEs, and they expand ≈96 % faster. Finally, while our results are generally consistent with previous studies, differences also arise due to (i) varying identification criteria; (ii) unequal ICME boundary times in catalogs; (iii) studies based on only subsets of ICMEs (e.g., sheaths or MCs); and (iv) differences in observation periods and datasets. These factors collectively highlight the difficulty in establishing uniform ICME statistics across missions and catalogs.

### *4.1 ICME Distribution during SC25 Rise Phase*

Figure 7 shows the yearly frequency of ICMEs in our catalog, separated into non-shocked events (slate blue) and shocked events (salmon rose) and overlaid with the annual mean sunspot number (SSN; green line). Note that all shock-associated ICMEs in our catalog contain a sheath region preceding the MC. The annual ICME count closely follows the sunspot number, starting low in 2020 and peaking in 2023–2024 before declining in 2025, with a positive correlation coefficient of ≈0.73. Previous catalogs, such as those from Wind and ACE spacecraft, have reported similar correlation coefficients (see Wu and R. P. Lepping 2011; Wu and Lepping 2016) over multiple solar cycles. When considered separately, both shocked and non-shocked ICME occurrences show significant correlations with solar activity. For instance, the correlation coefficients (CC) for non-shocked and shocked ICMEs are 0.60 and 0.69, respectively, indicating that both ICME groups closely track the SSN. However, it can be observed that the non-shocked ICMEs reached their peak in 2023 before the shocked ICMEs and SSN peaked in 2024. The predominance of shocked ICMEs during peak activity aligns with findings that faster CMEs (>500 km $s^{-1}$) are more likely to drive shocks, as the ICME speed exceeds the local magnetosonic speed (Gopalswamy et al. 2010).

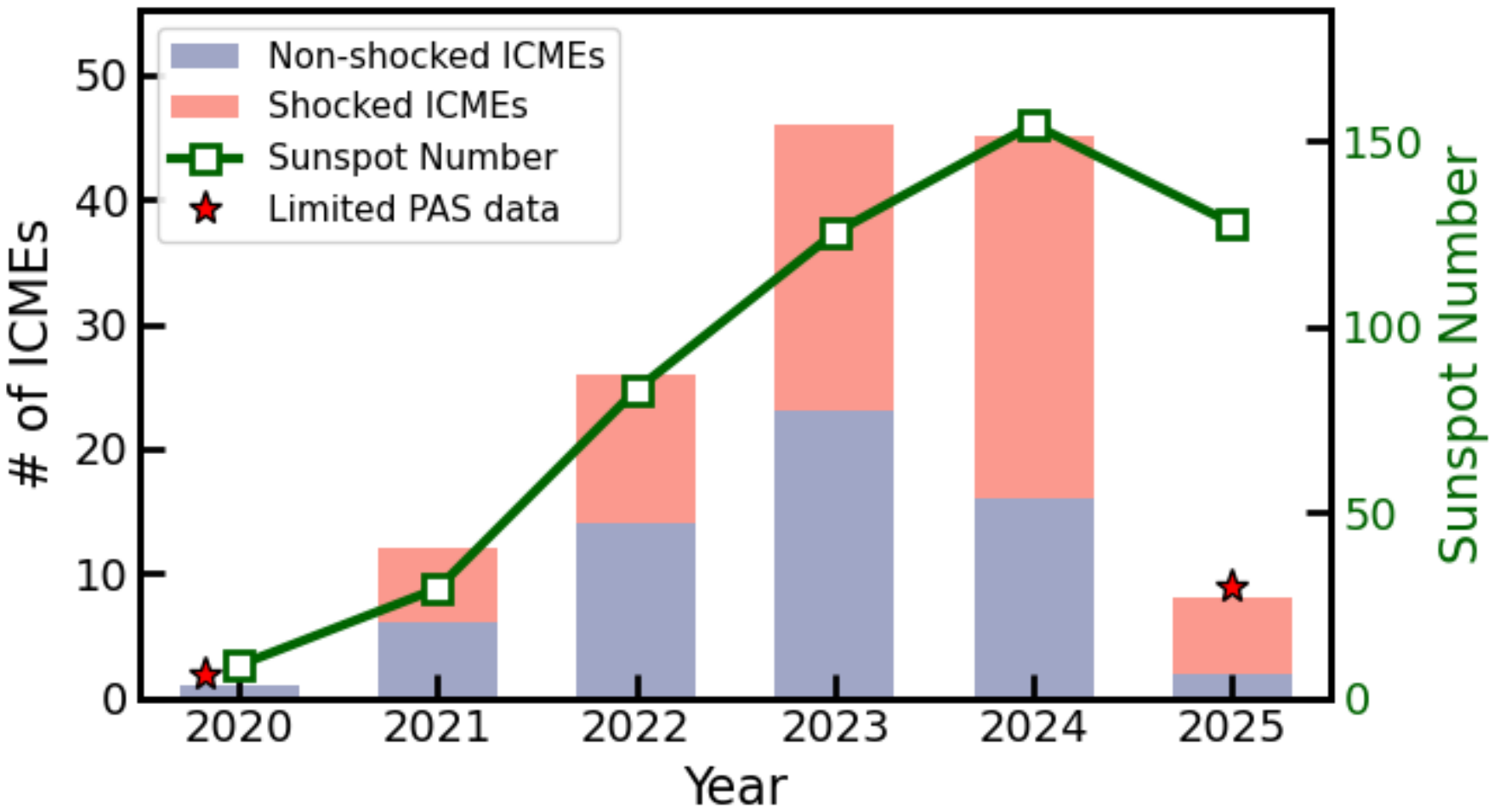

Figure 7. Shows the yearly frequency of ICMEs with shock (salmon rose), and without shock (slate blue), along with the annual mean sunspot number (green, right axis). * Indicates limited data.

### *4.2 Statistics of ICME Speed Parameters*

Figure 8 shows a nearly linear relationship (CC = 0.911)[4] between the shock and MC speeds. This strong correlation is in good agreement with previous studies by Gopalswamy et al. (2005), and Gopalswamy (2006) who obtained somewhat stronger correlations (CC= 0.93) and (CC = 0.95), respectively, suggesting that the tight coupling between shock and MC speeds is a significant feature across different event samples and/or methodologies. The dashed gray line marks where the shock and MC have equal speeds, corresponding to an approximate balance between the driving of the shock by the ejecta and the aerodynamic drag of the ambient solar wind. Events above this line indicate shocks that still outrun the ejecta, whereas events below it suggest that drag from the interplanetary medium has slowed the shock more strongly than the driver (see McComas et al. 2013). Additionally, it is not surprising that sheath proton temperature tends to increase systematically with both speeds, as indicated by the color bar. This implies that faster MCs tend to drive proportionally faster shocks and that stronger shocks produce hotter, more compressed sheaths through shock dissipation processes. The regression slope of 1.095 indicates that shock speeds in the inner heliosphere are, on average, ≈9.5% higher than bulk MC speeds, consistent with a piston-driven forward shock that still outruns the ejecta but remains dynamically coupled to it. This agrees well with Gopalswamy et al. (2015) who found that the average shock (sheath) speeds exceed the MC speeds by ≈10 % at 1 au during solar cycles 23 and 24.

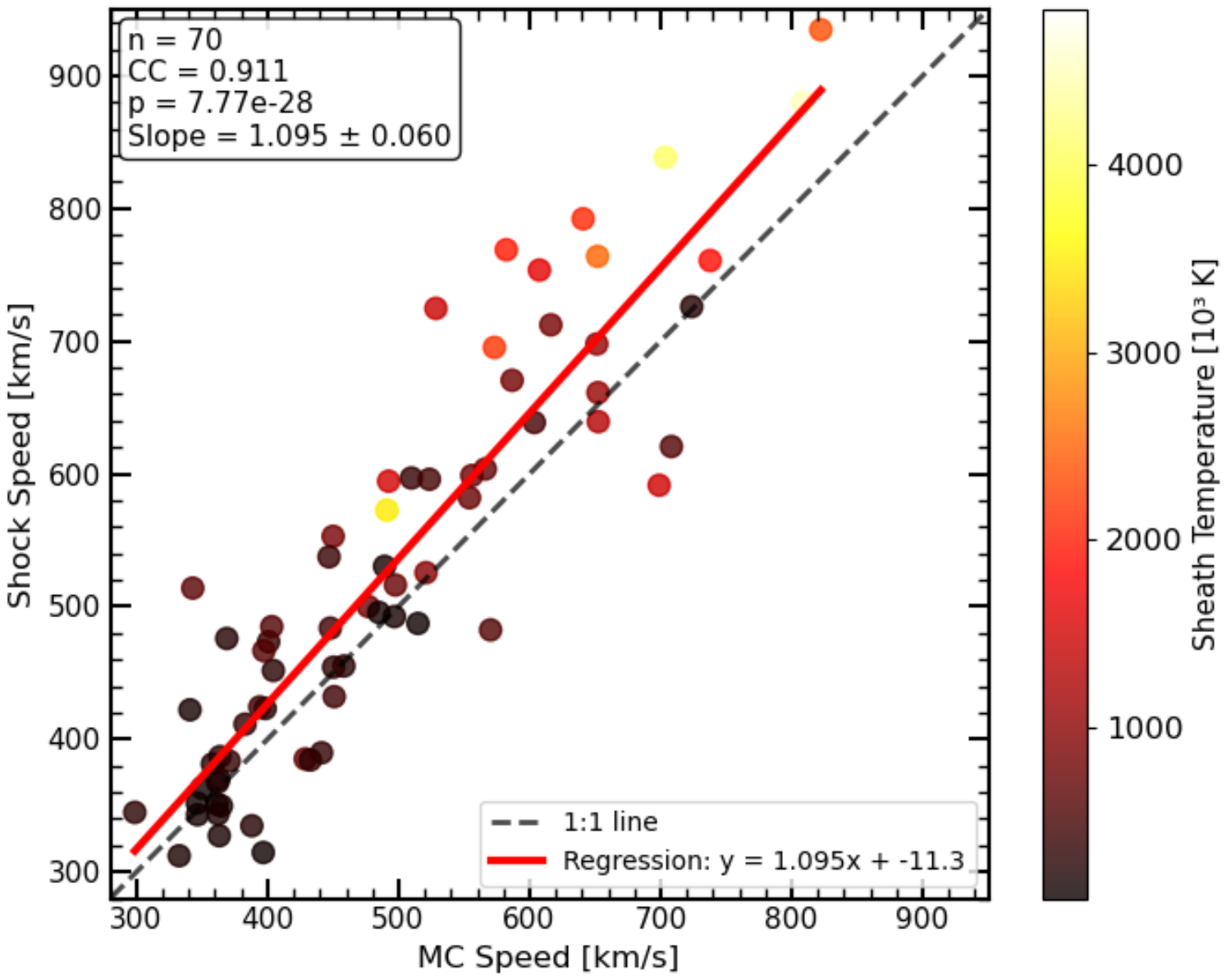

Figure 8. Shows the relationship between MC speeds and their associated shock speeds, color-coded by the sheath temperature. The dashed line denotes the 1:1 relation, while the solid red line shows the linear regression fit.

[4] Note: All correlation coefficients reported in this study are Pearson correlations.

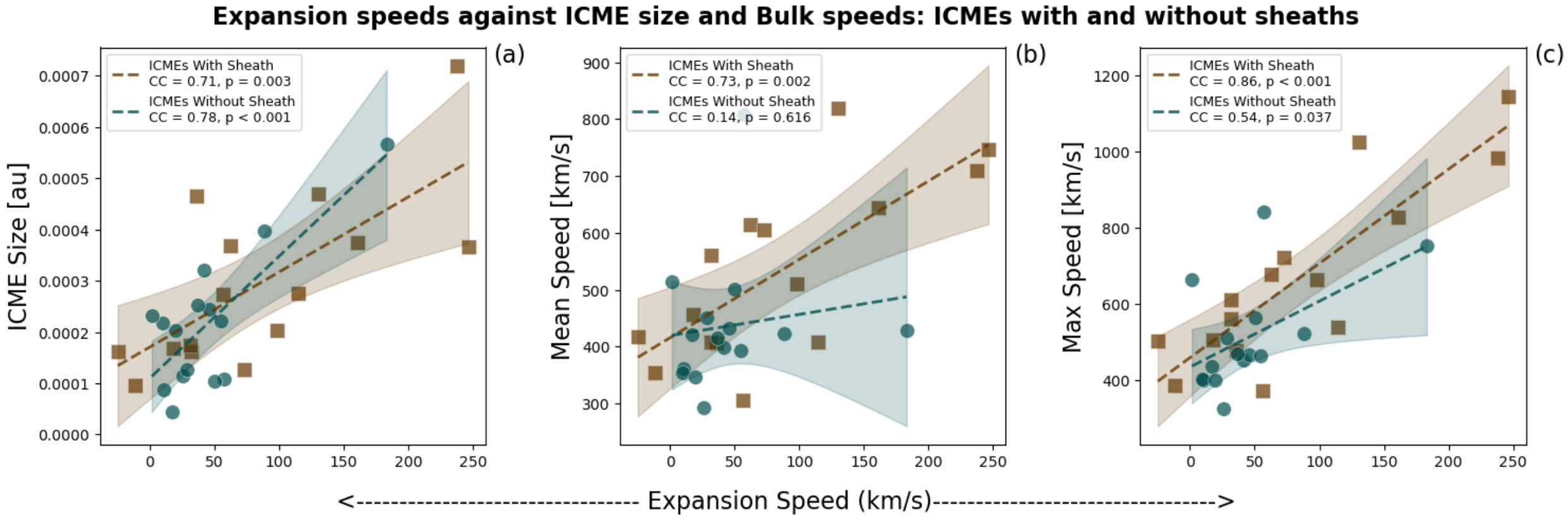

Figure 9. Show the relationship between ICME expansion speed (x-axis, km s⁻¹) and ICME size (a), mean ICME speed (b), and maximum ICME speed (c) for ICMEs with sheath (green) and without sheath (brown). The shaded bands denote 95% confidence intervals, while the legends contain the correlation coefficients and p-values in each panel.

At this point, we assess the relationship between ICME expansion speed and ICME radial size (Figure 9(a)), mean bulk speed (Figure 9(b)), and maximum bulk speed (Figure 9 (c)) for events with and without preceding sheath regions (Figure 9). We found strong positive correlations between ICME expansion speed and radial size, with CC = 0.708 ($p < 0.01$) for ICMEs with sheaths and a slightly stronger CC = 0.780 ($p < 0.001$) for those without sheaths. This confirms that ICME's larger radial sizes are tightly linked to greater internal expansion regardless of shock compression, consistent with previous statistical studies that expansion speed scales with ICME size (e.g., Gopalswamy et al. 2014), at expansion rates $\zeta \approx 0.6$-0.8 (Gulisano et al. 2010; Lugaz et al. 2020a; Mishra et al. 2021). Similarly, ICMEs with sheaths exhibit very strong, significant correlations between expansion speed and both mean bulk speed (CC = 0.730, $p < 0.01$) and especially maximum bulk speed (CC = 0.861, $p < 0.001$). These values provide quantitative evidence of discussions on in-situ kinematic relationships in recent studies (e.g., Salman et al. 2024; Larrodera and Temmer 2024; Trotta et al. 2024), indicating that fast-propagating ICMEs with shocks display stronger coupling between expansion and speed than non-sheath events, due to compression and turbulent heating in the sheath that sustains over-pressurized ejecta (Goslinag et al. 1994; Lugaz et al. 2020a). The particularly tight correlation with maximum speed further supports the interpretation that peak velocities within the ICME reflect the leading-edge dynamics most influenced by the upstream shock and pile-up region (Kilpua et al. 2017b; Manchester et al. 2017).

In contrast, ICMEs without sheaths show a dramatically weaker relationship with mean bulk speed (CC = 0.141, statistically insignificant) and only a moderate correlation with maximum speed (CC = 0.542, $p < 0.05$). This decoupling from mean speed aligns with prior results: non-shock ICMEs from streamer blowouts or quiescent filaments travel near slow solar wind speeds and expand mainly via residual magnetic pressure rather than shock forcing (see Liu et al. 2006; Manchester et al. 2017; Nitta et al. 2021). Non-sheath ICMEs nevertheless retain moderate maximum-speed correlation. This could mean localized high-speed fronts when flux ropes remain over-pressured relative to the ambient. Arguably, the presence or absence of a shock or sheath might fundamentally provide oppositely related mechanisms of the expansion–propagation relationship of ICMEs at 1 au. For instance, sheathed events, which are typically faster and potentially geoeffective (e.g., Owens et al. 2005, 2018), maintain strong kinematic coupling throughout their structure, whereas slower non-sheath ICMEs originate from cooler, less explosive source regions, expand independently of their bulk motion. Therefore, the results provide a valuable diagnostic for linking in-situ ICME properties to solar eruption characteristics and highlight sheath-driven dynamics in size–speed relationships.

As mentioned in Section 2, the HIS suite operated for roughly 16 months (January 2022–April 2023), during which 32 ICMEs contained composition measurements. Because heavy-ion coverage is limited, composition criteria are not incorporated into the automatic detection stage; instead, they are evaluated only after an ICME/MC has been identified. For each event, we compute the following standard composition diagnostics for further analysis: $\langle Q_{Fe} \rangle > 11$ (Gopalswamy et al. 2013; Lepri et al. 2001; Lepri & Zurbuchen 2004), $O^{7+}/O^{6+} < 0.6$ (Gopalswamy et al. 2013), ${F_e}^{\geq 16+}/F_e \geq 0.1$ (Lepri et al. 2001; Zurbuchen & Richardson 2006) and $F_e/O_{abun} > 2(F_e/O)_{pho}$ (Aellig et al. 1999; Asplund et al. 2021; Livi et al. 2023). Here, we adopt the nominal photospheric $(F_e/O)_{pho}$ value of 0.0589 (Asplund et al. 2021; Livi et al. 2023), since enhancements above twice this level reliably distinguish ICME plasma from the ambient wind during solar minimum. For instance, depending on the phase of a particular solar cycle, typical solar wind values display a low FIP bias ≈2–3× (Aellig et al. 1999), while ICMEs routinely show 2–4 times greater enhancements, and in rare cases they can exceed 5 times in flare-driven or SEP-contaminated conditions.

In Figure 10, we show the number and percentages of ICMEs satisfying each of the four compositional thresholds. Of the 32 ICMEs identified, only 28.1% exceeded $O^{7+}/O^{6+} < 0.6$, and tightening the threshold to >1, as suggested by Zurbuchen and Richardson (2006), reduced this to just 4 events. Though typical $O^{7+}/O^{6+}$ratios are ≈1, Richardson and Cane (2004b) also noted that the majority of their events have ratios between 0.2 and 8. Such solar wind streams are known to exhibit substantially low $O^{7+}/O^{6+}$ratios, and reduced $Fe$ charge states compared with slow wind (e.g., Von Steiger et al. 2000; Zurbuchen et al. 2002; Lepri et al. 2013). These compositional signatures reflect distinct freeze-in conditions and source-region histories and are consistent with the predominance of fast-wind structures during the ascending phase of SC25. Therefore, our result highlights that elevated $O$ charge states are not universal ICME identifiers. Instead, they are strongly influenced by coronal structures and electron thermal conditions at the eruption site in the lower corona (Gopalswamy et al. 2013; Lepri et al. 2013; Livi et al. 2023; Richardson & Cane 2004b).

On the contrary, we found $Fe$-group diagnostics, which are good proxies for electron temperature in the far corona (e.g., Lepri et al. 2001; Lepri and Zurbuchen 2004), flag a larger fraction of our ICMEs than the $O$-charge state criterion. In principle. ($\langle Q_{Fe} \rangle > 11$), ${F_e}^{\geq 16+}/F_e \geq 0.1$ and $F_e/O > 2(F_e/O)_{pho}$ have accounted for 62.5%, 50%, and 65.6% ICMEs, respectively, suggesting that majority are substantially filled with hotter plasma or originate from hotter source regions, consistent with the notion that stronger FIP-effect accompanies high charge states (see Zurbuchen et al. 2016). Comparing these values to previous studies, we noticed that the apparent differences with some earlier studies are largely methodological. For instance, Gopalswamy et al. (2013) identified 54 ICMEs (23 MC and 31 non-MC) solely based on $O$ and $\langle Q_{Fe} \rangle$ ion charge state and found that ≈95% MCs and ≈89 non-cloud ICMEs had peak $\langle Q_{Fe} \rangle \geq 11$. Similarly, Lepri et al. (2001) showed that long-duration ICMEs with shorter intervals of high charge states will miss strict detection criteria. In practice, they applied a 20-hour minimum duration to ≈90% of ICMEs identified using $F_e^{\geq 16+}/F_e \geq 0.1$, and obtained a significant reduction to ≈ 50% of ICMEs (i.e., consistent with our results). Richardson and Cane (2004b) also found that ≈72% ICMEs are associated with $F_e^{\geq 16+}$. Lepri and Zurbuchen (2004) revisited the problem to examine 57 ICMEs (duration < 6 hr.) and reported that ≈75% (as against 62.5% in this work) exhibited elevated $\langle Q_{\mathrm{Fe}} \rangle$ >12. Similarly, Song et al. (2016) also classified MCs by $\langle Q_{Fe} \rangle$, and inferred that ≈45%–50% have high-charge states, which is consistent with the percentages obtained in this study. Therefore, we suggest that by construction, 50–65% ICMEs showing $\langle Q_{Fe} \rangle$ or $F_e^{\geq 16+}/F_e$ enhancements are characteristically "hot ICMEs" from source regions with high coronal electron temperatures.

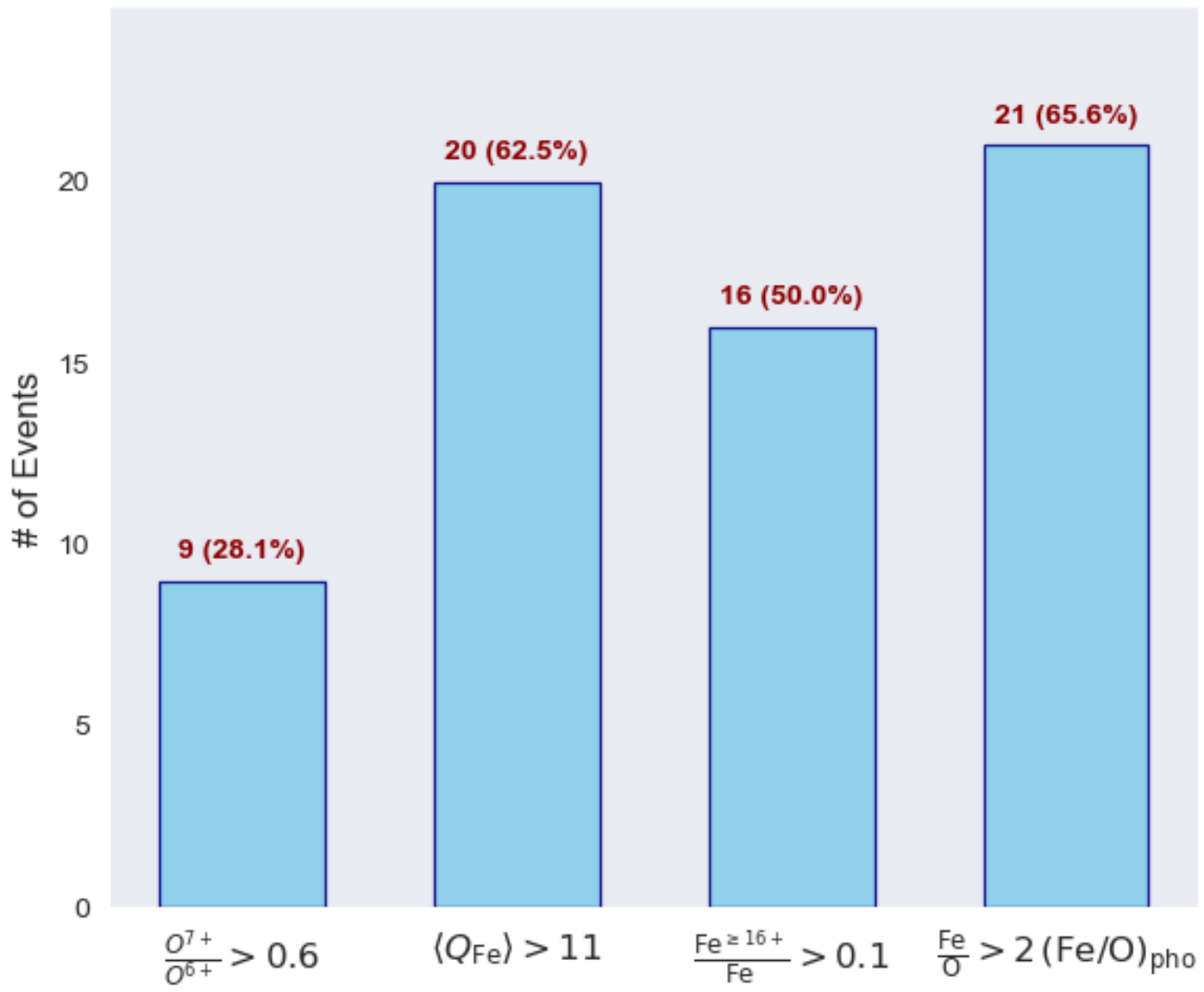

Figure 10. Distribution of 32 subset ICMEs (202-2023) satisfying composition metrics for identification. Bars show the number of ICMEs meeting each metric. Number of qualifying events and their percentage relative to the total ICMEs are indicated in red on each bar.

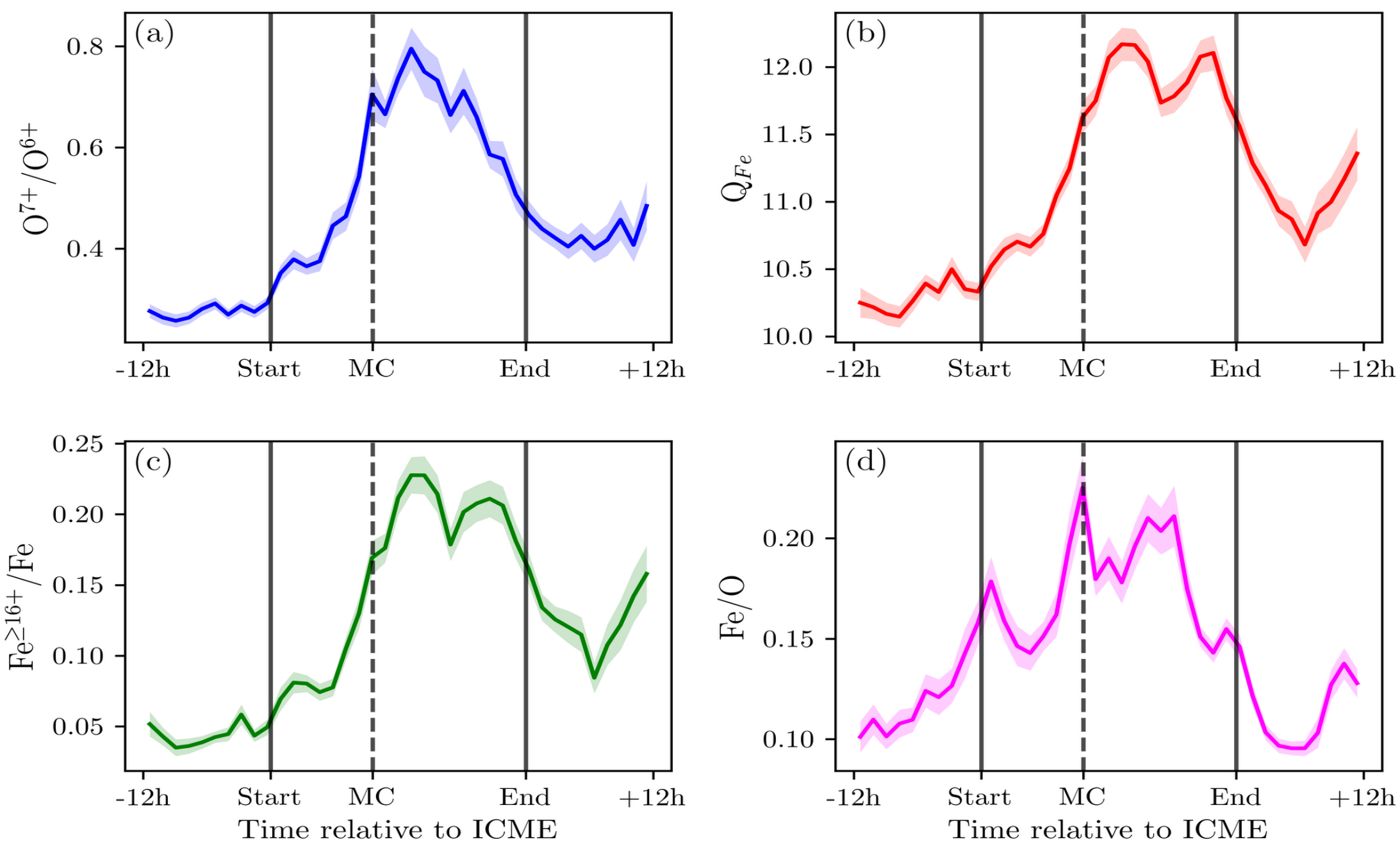

Figure 11: Shows superposed-epoch analysis of heavy-ion composition for Solar Orbiter ICMEs with HIS measurements. Composite times are the ICME start and end (solid vertical lines), and the MC onset (dashed vertical line). The analysis window extends from 12 h before the ICME start to 12 h after the ICME end. Curves show the mean profiles (solid lines) and associated 1-σ ranges (shaded) for (a) $O^{7+}/O^{6+}$, (b) $\langle Q_{Fe} \rangle$, (c) $Fe^{\geq 16+}/Fe$, and (d) *Fe/O.*

To emphasize the reliability of the composition criteria used in sorting out the 32 ICMEs from our catalog, we now consider the temporal profiles before, during, and after the ICMEs. Figure 11 (a-d) shows the superposed-epoch analysis of the heavy-ion composition for the 32 ICMEs as a function of normalized ICME time, and negative values correspond to the immediate upstream solar wind. In all four diagnostics, the upstream values remain near background levels, followed by clear enhancements shortly after the ICME onset and gradual decline toward the trailing edge. These profiles are very much consistent with those presented by Owens (2018). As expected, enhancements are most pronounced during the MC interval, indicating that the heavy-ion signatures are tightly associated with the core flux-rope plasma rather than the sheath or the immediate upstream solar wind.

In Figure 11(a), $O^{7+}/O^{6+}$ ratio exhibits a rapid increase near the ICME leading edge, reaching its maximum in the first half of the ejecta before progressively relaxing toward near-ambient values closer to the ICME end. This behavior is mirrored by $\langle Q_{Fe} \rangle$, which shows a strong rise above 11–12 in the ICME body, again with little to no enhancement in the pre-ICME solar wind (see Figure 11(b)). However, the fact that *O* and *Fe* charge states decay toward the trailing edge (but remain relatively higher at the leading edge) further suggests either radial mixing with cooler material or a temporal variation in the source region temperatures during CME eruption. The elevated $Fe^{\geq 16+}/Fe$ (Figure 11(c)) and *Fe/O* (Figure 11(d)) indicate strong *Fe* enhancement and significant FIP bias within the ICME body. Also, the *Fe* indicator groups display two peaks or plateau-like structures, which arguably may reflect internal substructures such as multiple flux ropes, erosion layers, or mixed plasmas from interacting CMEs. Again, a significant dip in the *Fe/O* profile can be observed in the sheath region and trailing edge of the ICME body, which is inconsistent with that shown in Figure 5h by Owens (2018). Overall, the ensemble profiles demonstrate that, on average, the automatically identified ICME intervals in our catalog indeed capture the bulk of the CME-processed material. Finally, we note that these results reinforce the interpretation that the heavy ion composition measurements by HIS provide an independent and physically motivated confirmation of the catalog, complementary to the magnetic and plasma criteria used in the detection algorithm. It also highlights the diagnostic power of composition for distinguishing ICME ejecta from the ambient solar wind and for constraining the coronal source conditions and heating histories of the events in our list.

## 5. Summary and Conclusion

We have automatically detected 138 ICMEs observed by Solar Orbiter using a distance-normalized and multi-threshold sliding window segmentation algorithm. We also quantify the radial evaluation of key solar-wind parameters observed that (i) proton temperature decreases with heliocentric distance more slowly than expected for adiabatic expansion, (ii) proton density follows a near-spherical radial decrease, (iii) dynamic pressure drops steeply, consistent with stronger ICME expansion in the inner heliosphere, and (iv) the magnetic-field radial dependence is shallower than the Parker-spiral approximation. As a way of validating the catalog, we used heavy ion composition and plasma metrics to show that, unlike non-sheath ICMEs, sheathed ICMEs contain elevated heavy ion charge states that are associated with faster expansion speeds, stronger sheath, and enhanced low–FIP element abundances. For ICMEs identified between 2020 – 2025, ≈70% are classified as flux-rope–like (Good or Complex MCs) and ≈30 % are nearly non-MC ICMEs. In addition, ≈55% were observed during the spacecraft's crossings, exhibiting the full shock–sheath–ejecta sequence, while the remaining 45% were observed at the flanks of the spacecraft.

From the yearly statistics, we showed that ICMEs with sheaths are systematically more extreme; on average, they have stronger fields, higher speeds, higher densities and dynamic pressures, and larger expansion speeds than non-sheath counterparts, while still maintaining low plasma $\beta$ within the ejecta. The yearly ICME counts track the rise of SC25 with a strong correlation to sunspot number, and the fraction of shock-driving events grows toward activity maximum, in line with the expectation that faster CMEs are more likely to produce interplanetary shocks. Within the subset where Solar Orbiter/HIS composition data are available (32 events, 2022–2023), the study finds that $O^{7+}/O^{6+}$ ratios are relatively unreliable indicators for ICME identification. In contrast, *Fe*-based diagnostics and *Fe/O* abundance serve as sensitive markers for sorting or classifying hot ICME structures. Thus, $O^{7+}/O^{6+}$ ratio tends to be more responsive to solar cycle dynamics than the *Fe* indicator groups. The superposed-epoch analysis further confirms that the heavy-ion indicators remain elevated in the ICME body, validating that our algorithm satisfactorily captures the ICME intervals listed in our catalog.

Finally, as a caveat, we highlight that the absence of an independent "ground truth" and methodological choices (signatures, thresholds, and matching rules) drive systematic differences in ICME event lists, providing a fundamental limitation in ICME research. Despite, the general consistency between the values reported in this study and that of previous studies, noticeable differences also arise due to several factors including (i) the use of varying identification criteria; (ii) unequal ICME boundary times in catalogs; (iii) studies based on only subsets of ICMEs (e.g. sheaths or MCs), (iv) and variations in observation periods and datasets. These factors collectively highlight the difficulty of establishing uniform ICME statistics across catalogs built based on observations from different spacecraft missions. Nonetheless, tying in-situ heavy-ion measurements and kinematic profiles back to plausible coronal conditions, this catalog and the methodology provide a framework for linking Solar Orbiter observations to the physics of CME initiation and offer a benchmark for future multi-mission ICME catalogs and for data-driven models of CME evolution and space weather impact.

## Acknowledgments

We thank the European Space Agency (ESA) for making the calibrated Solar Orbiter MAG, PAS, and HIS data publicly available. We also acknowledge the use of the ICMECAT and LASSOS catalogs for event comparison and validation. We are grateful to the anonymous reviewer for the valuable comments that significantly improved the quality of this manuscript.

## ORCID IDs:

Felix N. Minta https://orcid.org/0000-0002-9875-7540
Stefano A. Livi https://orcid.org/0000-0002-4149-7311
George C. Ho https://orcid.org/0000-0003-1093-2066
Susan T. Lepri https://orcid.org/0000-0003-1611-227X
Heather A. Elliott https://orcid.org/0000-0003-2297-3922

## Appendix A: Summary of the ICME Catalog

Appendix A is a summary of the initial version of the ICME catalog that was made in this study. For each event, the table lists magnetic-field properties, plasma moments, expansion metrics, sheath characteristics, and heavy-ion composition diagnostics.

| Column Number | Column Name | Description |
|---|---|---|
| 1 | ICME ID | Unique identifier for event. Constructed as SO_ICME-YYMMDDHHMM using the ICME/MC start time |
| 2 | ICME Start Time [UT] | Start of ICME interval (if a sheath was detected, this is the shock time; otherwise equals MC start). Formatted as YYYY-MM-DD HH:MM:SS |
| 3 | MC Start Time [UT] | Start time of the magnetic cloud / rotation interval. Formatted as YYYY-MM-DD HH:MM:SS |
| 4 | ICME End Time [UT] | End of ICME interval if MC ended before ICME. Formatted as YYYY-MM-DD HH:MM:SS |
| 5 | MC End Time [UT] | End time of the magnetic cloud / rotation interval. Formatted as YYYY-MM-DD HH:MM:SS |
| 6 | ICME Duration [hours] | Duration from ICME start (shock if present) to MC end in hours. |
| 7 | MC duration [hours] | Duration of the magnetic cloud hours |
| 8 | Mean \|B\| [nT] | Mean magnetic field during ICME/MC interval |
| 9 | Mean Br [nT] | Mean radial component during ICME/MC interval |
| 10 | Mean Bt [nT] | Mean tangential component during ICME/MC interval |
| 11 | Mean Bn [nT] | Mean normal component during ICME/MC interval |
| 12 | Mean Speed [km $s^{-1}$] | Mean bulk speed during ICME/MC interval. |
| 13 | Mean Temp [K] | Mean proton temperature over ICME/MC interval |
| 14 | Mean Density [$cm^{-3}$] | Mean proton density over ICME/MC interval |
| 15 | Mean Beta | Mean plasma $\beta$ over ICME interval |
| 16 | Vexp [km $s^{-1}$] | Expansion speed estimate computed inside MC: half the difference between max speed near the start and min speed near the end of the MC |
| 17 | Max \|B\| [nT] | Max magnetic field during ICME/MC interval |
| 18 | Max Speed [km $s^{-1}$] | Maximum bulk speed within ICME/MC |
| 19 | Max Temp [K] | Maximum proton temperature within ICME/MC |
| 20 | Max Density [$cm^{-3}$] | Maximum proton density within ICME/MC |
| 21 | Max Beta | Maximum plasma $\beta$ within ICME/MC |
| 22 | Min Beta | Minimum plasma $\beta$ within ICME/MC |
| 23 | Mean dyn pressure [nPa] | Mean dynamic pressure over ICME/MC |
| 24 | Max dyn pressure [nPa] | Maximum dynamic pressure over ICME/MC |
| 25 | sheath_duration [hours] | Duration between shock time and MC start time |
| 26 | Mean_B_sheath [nT] | Mean magnetic field in the sheath interval |
| 27 | Max_B_sheath [nT] | Maximum B in sheath |
| 28 | Mean_speed_sheath [km $s^{-1}$] | Mean bulk speed in the sheath |
| 29 | Max_speed_sheath [km $s^{-1}$] | Maximum bulk speed in the sheath |
| 30 | Min_speed_sheath [km $s^{-1}$] | Minimum bulk speed in the sheath |
| 31 | Mean_Temp_sheath [K] | Mean proton temperature in the sheath |
| 32 | Max_Temp_sheath [K] | Maximum proton temperature in sheath |
| 33 | Mean_density_sheath [$cm^{-3}$] | Mean proton density in sheath |
| 34 | Max_density_sheath [cm⁻³] | Maximum proton density in sheath |
| 35 | Mean_Beta_sheath | Mean plasma β in sheath |
| 36 | Max_beta_sheath | Maximum plasma β in sheath |
| 37 | Mean_Pdyn_sheath [nPa] | Mean dynamic pressure in sheath |
| 38 | Max_Pdyn_sheath [nPa] | Maximum dynamic pressure in sheath |
| 39 | Sheath Present? | Flag Y / N. Set to Y when a shock detected otherwise N. |
| 40 | Mean $O^{7+}/O^{6+}$ | Mean $O^{7+}/O^{6+}$ ionic ratio inside ICME/MC interval |
| 41 | $O^{7+}/O^{6+} >1$ | Logical flag (Y/N) indicating whether mean $O^{7+}/O^{6+}$ |
| 42 | Mean $Fe^{\geq 16+}/Fe$ | Mean fraction of *Fe* ions with charge ≥+16 divided by total Fe (ion composition diagnostic) |
| 43 | $Fe^{\geq 16+}/Fe \geq 0.1$ | Logical Flag (Y/N) whether Mean $Fe^{\geq 16+}/Fe \geq 0.1$. Set to Y when criterion is met. Otherwise set to N. |
| 44 | Mean $\langle Q_{Fe} \rangle$ | Mean iron charge state (average *Fe* charge state) within the ICME/MC |
| 45 | $\langle Q_{Fe} \rangle >12$ | Logical Flag (Y/N): Y if Mean $\langle Q_{Fe} \rangle > 12$ is satisfied, otherwise N. |
| 46 | Mean *Fe/O* | Mean *Fe/O* elemental abundance ratio inside ICME/MC |
| 47 | $Fe/O > 2 \cdot FeO_{pho}$ | Flag (Y/N) whether Mean *Fe/O* exceeds twice photospheric *Fe/O*. |
| 48 | Type | Automated tag assignment based on ICME/MC type. |
| 49 | Comments | Additional comments on the status of data availability or complexity of event. |
| 50 | RSO | Mean radial distance of solar orbiter during ICME/MC passage. |